\documentclass[prl,aps,superscriptaddress,twocolumn,longbibliography]{revtex4-2}
\usepackage{graphicx}
\usepackage{bm}
\usepackage{amsmath}
\usepackage{amsfonts}
\usepackage{amssymb}
\usepackage{epsfig}
\usepackage{color}
\usepackage{ulem}

\begin{document}

\title{Particle-hole spectral asymmetry at the edge of multi-orbital\\ noncentrosymmetric superconductors}

\author{Yuri Fukaya}
\affiliation{CNR-SPIN, I-84084 Fisciano (Salerno), Italy, c/o Universit\'a di Salerno, I-84084 Fisciano (Salerno), Italy}

\author{Keiji Yada}
\affiliation{Department of Applied Physics, Nagoya University, Nagoya 464-8603, Japan}

\author{Yukio Tanaka}
\affiliation{Department of Applied Physics, Nagoya University, Nagoya 464-8603, Japan}
\author{Paola Gentile}
\affiliation{CNR-SPIN, I-84084 Fisciano (Salerno), Italy, c/o Universit\'a di Salerno, I-84084 Fisciano (Salerno), Italy}

\author{Mario Cuoco}
\affiliation{CNR-SPIN, I-84084 Fisciano (Salerno), Italy, c/o Universit\'a di Salerno, I-84084 Fisciano (Salerno), Italy}

\begin{abstract}
Bogoliubov quasiparticles are a coherent electron-hole quantum superposition which typically, for time-reversal symmetric superconductors, exhibit a spectral distribution with particle-hole symmetry.
Here, we demonstrate that in two-dimensional noncentrosymmetric superconductors with multiorbital spin-triplet pairing the energy profile of the density of states at the edge can violate this paradigm. 
{{We show that the structure of Andreev reflections generally leads to pairing states made of configurations that are orbitally split due to the low degree of crystalline symmetry at the edge.
The resulting pairing state has a mixed parity character in the orbital sector that, in the presence of reduced crystal symmetry at the edge, sets out a particle-hole asymmetric profile for the spectral function. 
These findings indicate a path to design asymmetric spectral functions at the edge of superconductors with orbital degrees of freedom and time-reversal symmetry. The emerging signatures can be exploited for the detection of spin-triplet pairing equipped with internal degrees of freedom.}} 
\end{abstract}

\maketitle

\textit{Introduction}.\
Most of the quantum coherent properties of superconductors get manifested when the superconducting state is varying in space, as for instance at the edge, nearby the surface, or in proximity to another system.
In spatially inhomogeneous regions, indeed, multiple Andreev reflections~\cite{Andreev_reflection} lead to the formation of Andreev bound states with characteristic excitation energies below the superconducting
gap~\cite{Andreev_reflection,Bruder1990,KashiwayaTanaka2000RepProgPhys,Sauls2018,Prada2020,tanaka1995PRL}. The observation of these states, e.g., through scanning tunneling microscopy (STM) or angle-resolved photoemission spectroscopy (ARPES), is crucial to assess the nature of the superconducting phase and of the underlying pairing symmetry~\cite{KashiwayaTanaka2000RepProgPhys,tanaka1995PRL}. This is particularly challenging and relevant in superconductors having topologically protected boundary modes that are robust with respect to local perturbations~\cite{Qi2011,tanaka1995PRL,sato2017,Flensberg2021}. An emblematic case in this context is provided by the Majorana zero energy modes, as their successful generation and control represent a fundamental milestone for achieving topologically protected quantum computation \cite{Nayak2008}.

Recently, STM investigations of the just discovered heavy-fermion superconductor UTe$_2$ ~\cite{RanScience2019,DAokiJPSJ2019,DAokiJofP:CondMatter2022} revealed an asymmetric line shape of the density of states at surface step edges with features that break the particle–hole symmetry expected for Bogoliubov quasiparticles~\cite{JiaoNature2020}. Moreover, the energy asymmetric peaks that are observed inside the superconducting gap exhibit a behavior that depends on the direction of the normal to the side surface of the step edge~\cite{JiaoNature2020}. 
It is known that the particle-hole asymmetry in the superconducting spectra can descend from the normal state~\cite{Tanuma1997,Tanuma1999,YangJofP:cond_matter2007} or due to pair breaking scattering~\cite{Zou2021arXiv}, as it occurs in cuprates due to their large energy gap. Similar asymmetric features are also expected in the presence of sources of time-reversal symmetry breaking~\cite{NgPRB2004,Linder2010,EschrigRPP2015,Lu_2015,Lu_2018,Ando2022} as well as due to mechanisms driven by electron-boson interactions~\cite{SetiawanPRR2021,Cuoco2004,Cuoco2006} and {{dissipative coupling}}~\cite{Woods2019,Liu2017}. 

{{In this context, a fundamental and general question arises: what type of time-reversal symmetric superconductor can accommodate Andreev bound states that lead to a particle-hole asymmetric profile of the density of states?}}
In this Letter, we provide an answer to this problem.
{{We demonstrate that in two-dimensional noncentrosymmetric superconductors with multiorbital $s$-wave spin-triplet pairing the density of states at the edge is generally marked by an asymmetric line shape. 
We show that the lack of horizontal mirror symmetry induces a distinct type of orbital-dependent Andreev reflections, via the so-called orbital Rashba interaction ~\cite{park11,park12,Khalsa2013PRB,kim14,Mercaldo2020}, that is able to set out a pairing at the edge with orbital mixed parity in the spin-triplet channel. This pairing state indeed includes configurations with orbital singlet ($L=0$) and orbital quintet ($L=2$) angular momentum (or equivalently mixed singlet-triplet configurations in the pseudospin manifold spanned by two orbitals). 
We demonstrate that this type of pairing in the presence of the orbitally driven interactions that reduce the crystal symmetry at the edge of the superconductor, can generally yield asymmetric spectral distribution of the density of states.}}

\begin{figure*}[htbp]
    \centering
    \includegraphics[width=17cm]{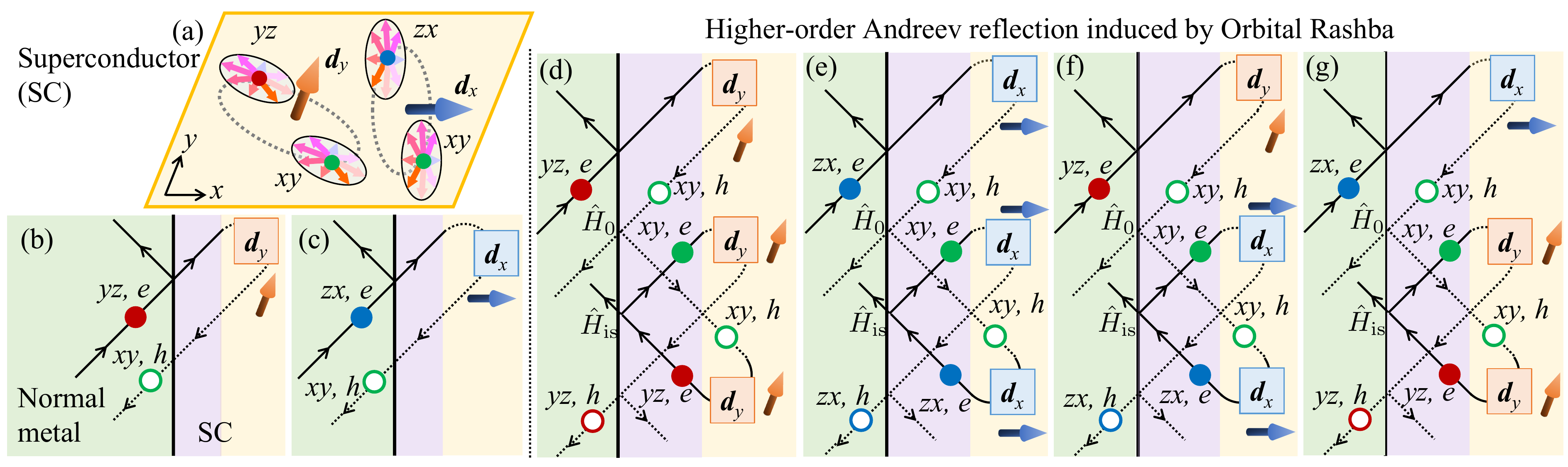}
    \caption{(a) Sketch of the interorbital spin-triplet/orbital-singlet/$s$-wave pairing (B$_1$ symmetry for the $C_{4v}$ crystal point group).
    The pairing is described by two \textbf{d} vectors. $\bm{d}_{y}$ for the $(xy,yz)$ orbital channels and $\bm{d}_{x}$ for the $(xy,zx)$ sector.
    (b),(c) The lowest-order Andreev reflection (AR) that originates from the interorbital spin-triplet pairing in (a) at the normal metal (N)/superconductor (S) interface.
    Due to the ${\bf d}$ vector structure, as allowed by the crystal symmetry, AR for $[(yz,e),(xy,h)]$ in (b) is not equivalent to that for $[(zx,e),(xy,h)]$ in (c).
    Here, $e$ ($h$) denotes the electron (hole) quasiparticles.
    (d)-(g) Higher-order Andreev reflection caused by OR for (d) $[(yz,e),(yz,h)]$, (e) $[(zx,e),(zx,h)]$, (f) $[(yz,e),(zx,h)]$, and {(g) $[(zx,e),(yz,h)]$} electronic states, respectively.
    $\hat{H}_{0}$ and $\hat{H}_\mathrm{is}$ indicate the scattering processes involving the hopping term and the OR coupling, respectively.
    Since the pair potential is associated to spin-triplet pairs, for injected electrons and reflected holes, the spin orientation is not modified, i.e., the relevant AR processes are spin conserving.}
    \label{fig:1}
\end{figure*}%
\begin{figure}[htbp]
    \centering
    \includegraphics[width=7.2cm]{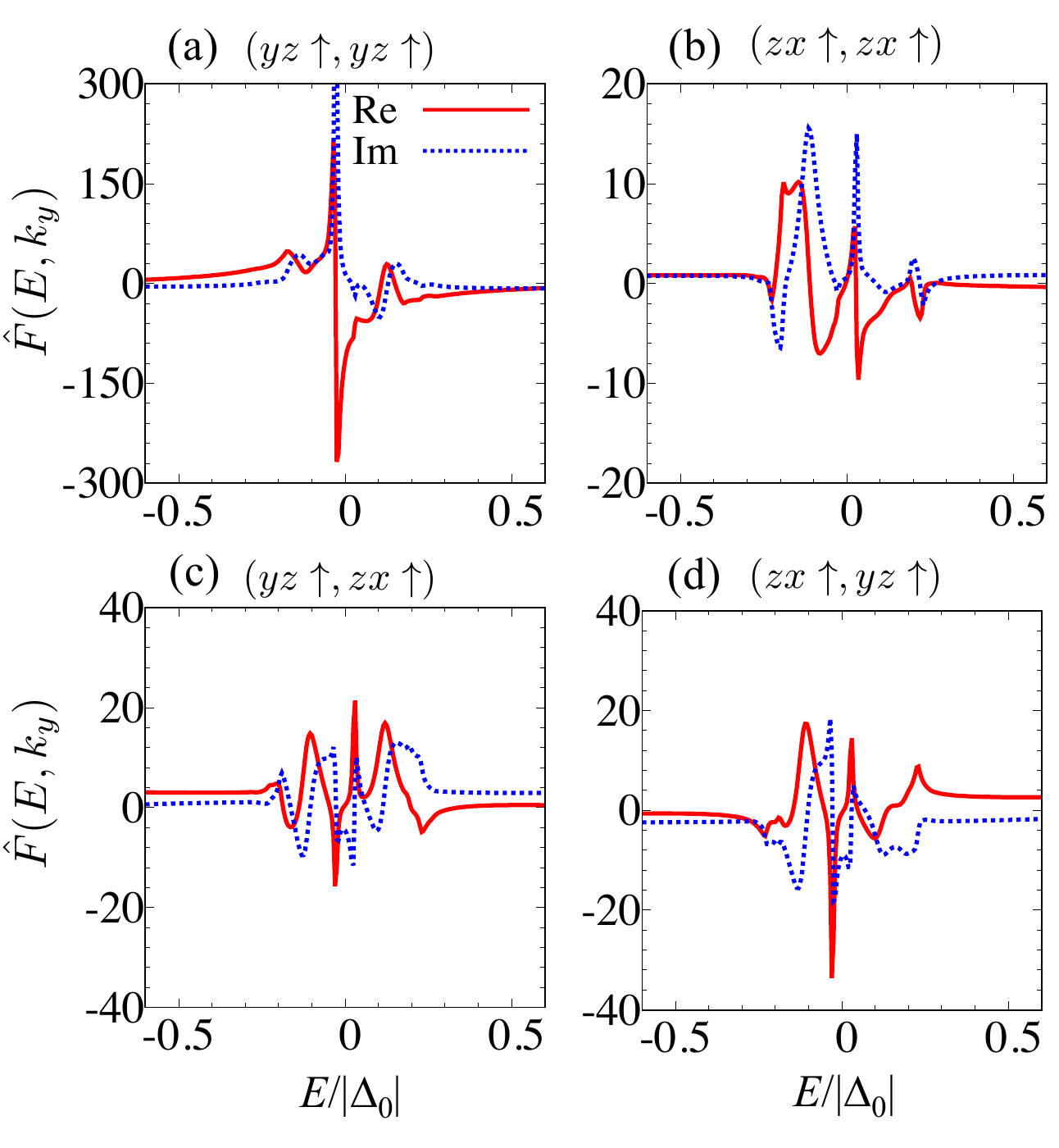}
    \caption{Anomalous part of the retarded Green's function $\hat{F}(E,k_y)$ {for the B$_1$ representation} at {a representative momentum $k_y=0.05\pi$} at the (100) termination for {the equal-spin intra-orbital channels (a) $(yz\uparrow,yz\uparrow)$ and (b) $(zx\uparrow,zx\uparrow)$, and equal-spin interorbital components (c) $(yz\uparrow,zx\uparrow)$ and (d) $(zx\uparrow,yz\uparrow)$}.
    Red-solid (blue-dotted) line denotes the real (imaginary) part of the anomalous part. 
    At finite momentum the real part of the anomalous Green's function has both even and odd frequency components.
    {The orbital Rashba interaction is $\alpha_\mathrm{OR}/t=0.20$, the chemical potential $\mu/t=0.35$, and the gap amplitude $|\Delta_{0}|=0.0010t$, respectively.}}
    \label{fig:2}
\end{figure}


\textit{Model}.\
To illustrate this physical scenario, we consider a multiorbital 2D electronic system with spin-triplet $s$-wave pairing and broken inversion symmetry.
The electronic structure is marked by three bands arising from atomic states spanning an $L=1$ angular momentum subspace, such as $d_a$ orbitals with $a=(yz,zx,xy)$. Here, we refer to $d$-orbitals localized at the site of a square lattice assuming a $C_{4v}$ point group symmetry. 
The breaking of planar mirror symmetry ($M_z$) sets out a polar axis, $z$, leading to an orbital Rashba interaction ($\alpha_\mathrm{OR}$) that mixes orbital configurations with different mirror parity, i.e., $xy$ with $xz$ or $yz$.
As for the spin Rashba interaction, the orbital Rashba couples the atomic angular momentum ${\bf L}$ with the crystal wave vector $\bm{k}$ in the usual form, i.e.\ $\alpha_\mathrm{OR} [\hat{L}_x \sin(k_y)-\hat{L}_y \sin(k_x)$]~\cite{park11,park12,Khalsa2013PRB,kim14,Mercaldo2020}. 
In the manifold of the $d_a$ orbitals, the components of the angular momentum can be expressed in a matrix form as $[\hat{L}_k]_{lm}=i \epsilon_{klm}$ with $\epsilon_{klm}$ the Levi-Civita tensor, while hereafter $\hat{s}_{j=x,y,z}$ refers to the Pauli matrix in spin space, with $\hat{s}_0$ being the identity matrix.
Assuming that the basis of the local creation operator of electrons for each $d$-orbital is
$\hat{C}^{\dagger}_{\bm{k}}=[c^{\dagger}_{yz,\uparrow\bm{k}},c^{\dagger}_{zx,\uparrow\bm{k}},c^{\dagger}_{xy,\uparrow\bm{k}},c^{\dagger}_{yz,\downarrow\bm{k}},c^{\dagger}_{zx,\downarrow\bm{k}},c^{\dagger}_{xy,\downarrow\bm{k}}]$, 
the Hamiltonian can be generally expressed as 
$\hat{\mathcal{H}}=\sum_{\bm{k}}\hat{C}^{\dagger}_{\bm{k}}\hat{H}(\bm{k})\hat{C}_{\bm{k}}$
with $\hat{\mathcal{H}}(\bm{k})$ given by
\begin{eqnarray*}
\label{Eq:h0}
\hat{\mathcal{H}}(\bm{k})&&=\sum_a [\epsilon_{a}(\bm{k})\mathbb{P}_a+\alpha_{OR} \left( {\bf g}_{\bm{k}}\times \hat{\bf L} \right) \cdot \hat{z}-\mu]\otimes \hat{s}_{0} \\ &&+ \lambda_\mathrm{SO}\hat{\bf{L}}\cdot\hat{\bm{s}},
\end{eqnarray*}
where ${\bf g}_{\bm{k}}=(\sin(k_x),\sin(k_y),0)$, 
while $\epsilon_a({\bm{k}})=-2t^x_a\cos(k_x)-2t^y_a\cos(k_y)-\delta_a$ indicate the dispersion relations for each orbital $a$. The nearest-neighbor hopping amplitudes are assumed to be $t_{yz}^y=t_{xz}^x=t_{xy}^x=t_{xy}^y=t$, $t_{yz}^x=t_{xz}^y=t'$, and $\mathbb{P}_a=(\hat{L}^2-\hat{L}_a^2)/2$ is a projector on the orbital $a$ subspace. At the $\Gamma$ point of the Brillouin zone, a degeneracy occurs between the $xz$ and the $yz$ orbitals, thus $\delta_{xz}=\delta_{yz}=0$. In turn, the orbital $xy$ has a nonvanishing crystal field splitting, $\delta_{xy}\equiv \Delta_t$, whose amplitude depends on the crystalline distortions. 
$\lambda_\mathrm{SO}$ is the atomic spin-orbit coupling expressing the interaction between the spin and angular momentum at each site.

One can tune the number of Fermi lines from two to six by varying the chemical potential $\mu$. Here, we focus on the configuration with six Fermi lines, by choosing a representative value for $\mu$, i.e.\ $\mu/t=0.35$, {{because we aim to have all the orbitals active at the Fermi level}}. Amplitude variations of $\mu$ that keep the number of Fermi lines do not alter the qualitative outcome of the results.
For the superconducting state, the Bogoliubov-de Gennes (BdG) Hamiltonian is then directly constructed by including the pair potential $\hat{\Delta}(\bm{k})$. 
{{
We consider different symmetry allowed local ($s$-wave) spin-triplet pairing with orbital-singlet character. The various irreducible representations can involve $s$-wave ($\bm{k}$-independent) ${\bf d}$ vector components for any given pair of orbitals~\cite{fukaya18}. The focus here is on superconducting states with B$_1$ and A$_1$ symmetry in the $C_{4v}$ group.}}
The B$_1$ order parameter is described~\cite{fukaya18,fukaya19,fukaya20} by an $s$-wave ${\bf d}$ vector with ${d}_x$ (${d}_y$) components corresponding to local electron pairs between $\{d_{zx},d_{xy}\}$ ($\{d_{zy},d_{xy}\}$) orbitals, respectively. We recall, that the ${d}_x$ and ${d}_y$ components refer to spin-triplet pairs having zero spin projections along the $x$ and $y$ directions, respectively. Due to the $C_{4v}$ symmetry, $d^{(xy,zx)}_{x}$ and $d^{(xy,yz)}_{y}$ are equal in amplitude. A schematic illustration of the spin and orbital structure of the B$_1$ Cooper pair is shown in Fig.~\ref{fig:1} (a). {{We notice that for this configuration the pair potential in the bulk does not involve all the orbitals as electrons in the $xz$ and $yz$ orbitals are not directly paired. Hence, it allows us to investigate the orbital reconstruction of the pairing at the edge.}}
The B$_1$ phase exhibits nodal points along the diagonal of the Brillouin zone ([110] direction) with nonzero topological number, due to the chiral symmetry of the BdG Hamiltonian ~\cite{Sato2011PRB,Yada2011PRB,Brydon2011PRB,Mercaldo16,fukaya18,fukaya19,fukaya20,Fukaya2022} and can lead to anomalous Josephson effects~\cite{fukaya20,Fukaya2022}.
{{The A$_1$ pairing configuration instead yields a fully gapped superconducting phase. Additionally, the A$_1$ $s$-wave ${\bf d}$ vector components can be non-vanishing for any given pair of orbitals in the ($zx$,$yz$,$xy$) manifold (see Supplemental Material \cite{SupMat}). The comparison of the A$_1$ and B$_1$ phases, hence, allows us to discern the role of topological nodal points and the orbital dependence of the ${\bf d}$ vector components. }}

\textit{Orbital dependent Andreev reflection and anomalous pair correlations}.\
Andreev~\cite{Andreev_reflection} considered the reflection of electrons incident from a normal region
into a superconducting domain of the same material as schematically described in Fig.\ \ref{fig:1}b. For energies below the gap, the electron is forbidden to propagate into the superconducting region and the scattering from the pair potential in the superconductor convert an electron into a hole. When dealing with the spin-triplet superconductor upon examination the Andreev reflection has an inherently richer spin-orbital structure [Fig.~\ref{fig:1} (b) and Fig.~\ref{fig:1} (c)]. Indeed, a spin polarized electron-like excitation with, for instance, $xz$ orbital character, upon reflection is converted into an outgoing $xy$ hole excitation with the same spin orientation. This reflection involves the ${d}_y$ spin-triplet pair potential [Fig.~\ref{fig:1} (b).] A similar process occurs for the $zx$ electron-like excitation that, via the ${d}_x$ pair potential, get converted into an outgoing $xy$ hole excitation [Fig.~\ref{fig:1} (c)]. We thus observe that basic Andreev reflections involve conversion among orbital states with opposite $M_z$ mirror parity. However, we note that the orbital Rashba coupling involves charge transfers along the $x$ and $y$ directions [Fig.\ \ref{fig:1} (d)] with orbital hybridization that does not conserve the $M_z$ mirror parity. This aspect is thus relevant to allow for high-order Andreev reflections that can convert electron into hole states with the same horizontal mirror parity. The resulting processes are schematically described in Figs.\ \ref{fig:1} (e) and \ref{fig:1} (f). Indeed, multiple reflections by the superconducting pair potential combined with intermediate orbital mixing, through the orbital-Rashba coupling ($H_\mathrm{is}$), leads to both $yz$-electron to $yz$-hole and $yz$-electron to $zx$-hole conversion. We observe that the inter-orbital $yz$-$zx$ reflection requires both $\bm{d}_x$ and $\bm{d}_y$ pair potentials; thus an effective rotation in the spin space of the spin-triplet ${\bf d}$ vector is needed. On the other hand, the equal orbital (i.e., $zx$ or $yz$) electron-hole conversion is achieved without rotating the ${\bf d}$ vector. These processes indicate that the induced intra- and interorbital pairing amplitudes are generally not equivalent, thus leading to an orbital mixed parity configuration. 

In order to verify and quantify the orbital structure of the pair correlations at the edge we evaluate the corresponding anomalous Green's function. The analysis is performed by means of the recursive Green's function method~\cite{Lee_FisherPRL1981,LDOSUmerski,KawaiPRB} assuming a semi-infinite geometry for the superconductor. The anomalous Green's function, $\hat{F}(E,k_y)$, is then evaluated for equal spin orientation, i.e.\ $\uparrow\uparrow$ and $\downarrow\downarrow$, and within the ($zx,yz$) orbital manifold, at the edge of the superconductor, assuming that the termination is perpendicular to the (100) direction. 
The results demonstrate that the edge pair correlations induced in the $(xz,yz)$ manifold for spin-triplet pairs have a rich orbital structure (Fig.~\ref{fig:2}). At zero momentum the real part of the anomalous Green's function is purely odd-in-frequency \cite{Tanaka2012,Tanaka2007, Tanaka2007a, Eschrig2007, Linder2019, Black2013, Asano2015,Ebisu2016}, while at finite momentum it exhibits both even and odd frequency components.
This is expected on the basis of the Fermi-Dirac statistics and the multi-orbital character of the paring amplitude. 
Moreover, we observe that, due to the matrix structure of $\hat{L}$, the eigenstates of its components $\hat{L}_{k}$, with non-vanishing quantum numbers $(+1,-1)$, are generally expressed as combinations of two out of three states with zero orbital polarization. Hence one has, for instance, that for $L_z$ the $(+1,-1)$ configurations are given by $|\pm \rangle_z = \frac{1}{\sqrt{2}} \left( \mp i |xz\rangle + |yz\rangle   \right)$.
Then, in the $(xz,yz)$ orbital manifold, since the intra- and inter-orbital anomalous Green functions are non-vanishing, one can have pairs with total orbital angular momentum $L_z=2$, i.e. with \ $+_z+_z$ and $-_z-_z$ configurations. Taking into account that for the spin channel the pairs have a spin-triplet structure, and by combining the spin and orbital angular components, $J_z=L_z+S_z$, one ends up with an effective pairing configuration with $J_z=3$ and thus a septet paired state. 
{{Moreover, since the inter-orbital pair amplitudes are not equivalent for exchange of the orbital index, the resulting state includes an orbital singlet configuration too. Such orbital singlet and quintet mixed parity state is a general mark of the pairing structure at the edge for all the pairs of orbital channels in the ($zx$,$yz$,$xy$) manifold.}}

\begin{figure}[htbp]
    \centering
    \includegraphics[width=7.1cm]{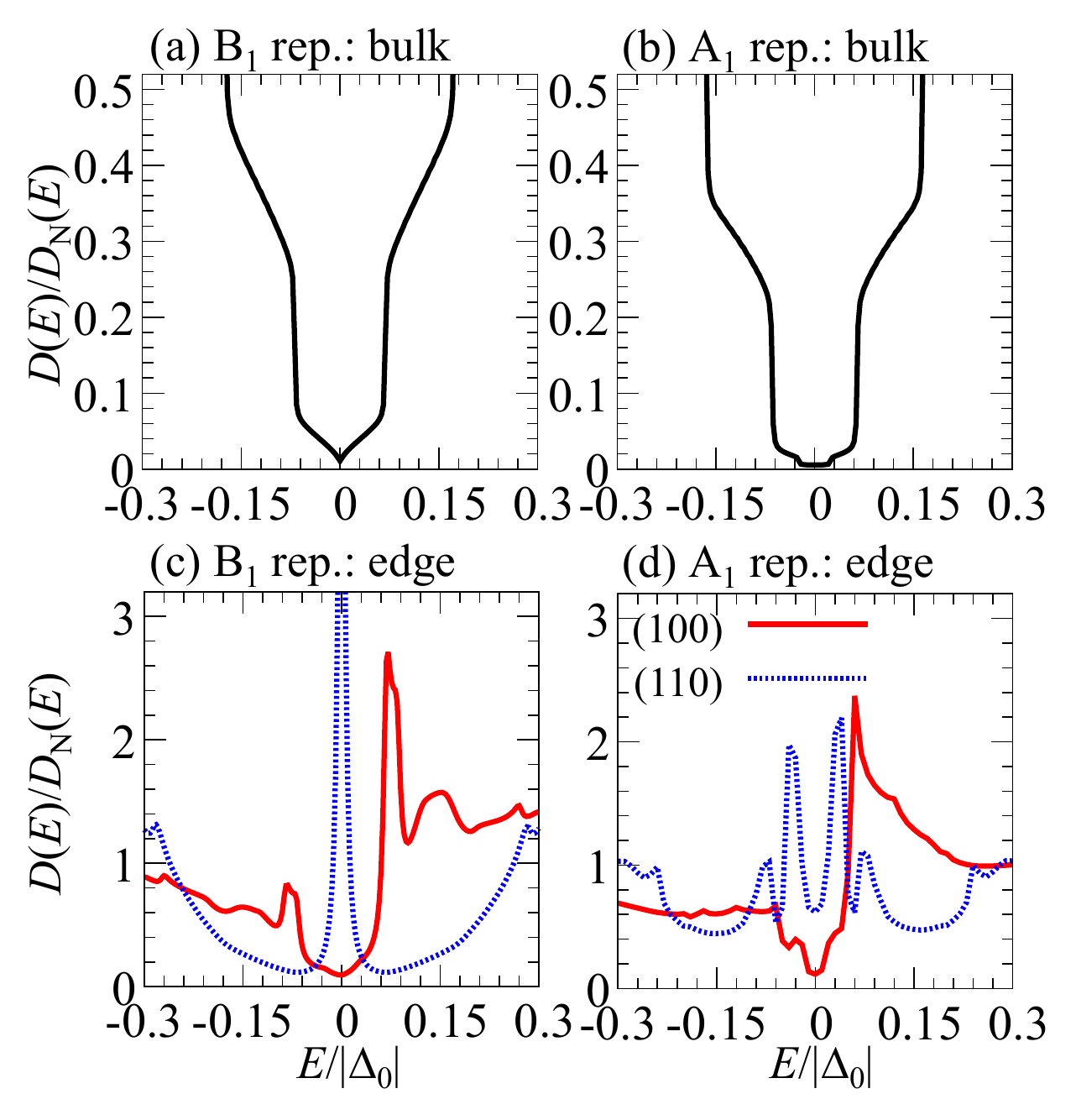}
    \caption{{Density of states (DOS) $D(E)$ in the bulk (a),(b) and at the edge (c),(d) for a representative configuration with 
    spin-triplet/orbital-singlet/$s$-wave B$_1$ $(d^{(xy,zx)}_{x},d^{(xy,yz)}_{y})=(|\Delta_0|,|\Delta_0|)$ and A$_1$ states $(d^{(xy,zx)}_{x},d^{(xy,yz)}_{y},d^{(yz,zx)}_{z})=(-|\Delta_0|,|\Delta_0|,0.0)$.
    We choose a representative orbital Rashba coupling as $\alpha_\mathrm{OR}/t=0.30$ while the chemical potential is given by  $\mu/t=0.35$. The other electronic parameters are $t'=0.01 t$, $\Delta_\mathrm{t}=0.5 t$, $\lambda_\mathrm{SO}=0.1 t$, and the gap amplitude $|\Delta_{0}|=0.001t$.}}
    \label{fig:3}
\end{figure}%
\textit{Asymmetric peaks in the density of states}.\ We then move to the analysis of the density of the states at the edge of the superconductor by considering a 2D geometry with different terminations [i.e., (100) and (110)] and for the B$_1$ and A$_1$ pairing symmetry in the bulk.
The density of states (DOS) $D(E)$ is calculated by the retarded Green's function $\tilde{G}(E,k_y)$,
\begin{align}
    D(E)&=-\frac{1}{\pi}\mathrm{Im}\int dk_y\mathrm{Tr}'[\tilde{G}(E,k_y)].
\end{align}%
In the superconducting state, the trace is performed by singling out the electron contribution.
We adopt the recursive Green's function method~\cite{LDOSUmerski} to obtain the retarded Green's function $\tilde{G}(E,k_y)$ at a surface layer (see Supplemental Material \cite{SupMat}).

As mentioned before, the bulk DOS for the B$_1$ [Fig. \ref{fig:3}(a)] and A$_1$ [Fig. \ref{fig:3}(b)] indicates that the B$_1$ is nodal while the A$_1$ has a small gap.
The evaluation of the DOS at the edge for both the B$_1$ [Fig. \ref{fig:3}(c)] and A$_1$ [Fig. \ref{fig:3}(d)] superconducting phases reveals an asymmetric line shape that occurs only for the (100) termination. For the A$_1$ the asymmetric profiles occur for any choice of the ${\bf d}$ vector structure (see Supplemental Material \cite{SupMat}).
The asymmetric spectral weight distribution has a significant orbital dependence among the $xz$ and $yz$ channels (see Supplemental Material~\cite{SupMat} for details). 
This aspect reflects the fact that $xz$ and $yz$ orbitals have a significant orbital anisotropy in the kinetic term along the (100) termination, thus resulting into a substantial rotational symmetry breaking at the boundary. 
On the other hand, for the (110) case, both $xz$ and $yz$ have a similar hopping amplitude along the (110) and the other diagonal (1$\hat{1}$0) direction with a low degree of orbital anisotropy and rotational symmetry breaking.
Since an asymmetric DOS is obtained for both the A$_1$ and B$_1$ phases, we also observe that the nodal structure of the gap function is not a needed feature for the achieved effects.

\begin{figure}[htbp]
    \centering
    \includegraphics[width=7.5cm]{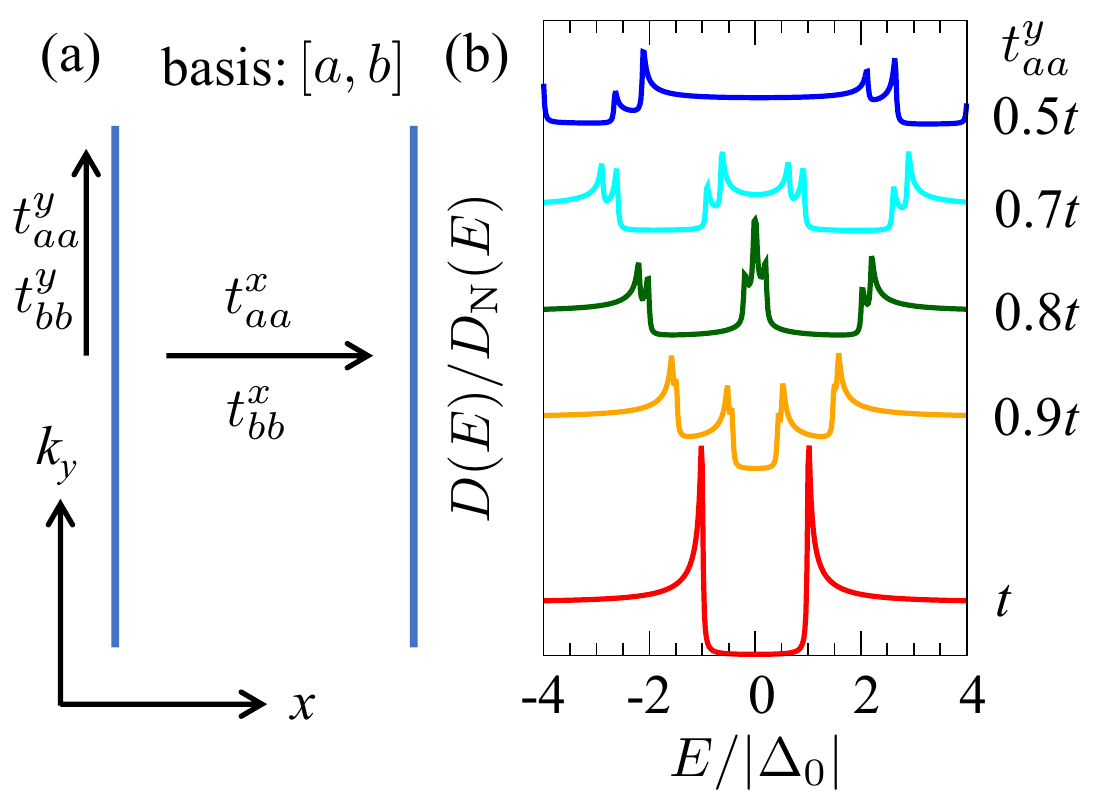}
    \caption{{(a) Schematic illustration of two superconducting chains in the $xy$ plane with different spatial position along $x$-axis and translational symmetry along $y$. $t_{\gamma\gamma'}^{i}$ with $\gamma,\gamma'=a,b$ and $i=x,y$ describe the effective hopping amplitudes within the ($xz$,$yz$) manifold along the $x$ and $y$ directions. (b) Local density of states for the ($xz$,$yz$) orbitals as a function of the $t_{aa}^{y}$ hopping parameter that controls the orbital anisotropy, while keeping $t_{bb}^{y}=t$ and the strength of the $C_4$ rotation symmetry in the finite size two-chains problem. The reduction of the $t_{aa}^{y}$ amplitude yields a splitting of the orbital excitations and an asymmetric profile of the density of states.}}
 
    \label{fig:4}
\end{figure}%

{\textit{Effective model and mechanism for asymmetric DOS.}}
At this point, we would like to provide a minimal model to capture the mechanisms related to the occurrence of the asymmetric density of states at the edge. To this aim, we consider an effective model with equal spin (i.e.\ spin-triplet) pairing and with only two-orbital degrees of freedom. We focus on the orbital manifold spanned by the ($xz$,$yz$) states which have opposite parity with respect to the vertical mirrors (e.g.\ $M_x$) and are energy degenerate in the tetragonal configuration with $C_{4v}$ point group symmetry. This implies that at the edge they get split and coupled by crystalline interactions that break mirror and rotation symmetries. 
Then, in order to make a direct connection with the problem upon examination that breaks translational symmetry and lacks rotation and vertical mirror symmetries at the boundary, we consider a model Hamiltonian in real space for a limited number of chains. Here, on the basis of the complete analysis of the anomalous correlator at the edge, we can assume that the pairing amplitude is given by an $s$-wave equal spin configuration (spin-triplet) with orbital-singlet. This is instead generally realized due to the orbital Andreev reflections and the reduced crystalline symmetry at the edge as demonstrated by the behavior of the anomalous Green's function in Fig.\ 2, even when in the bulk the $s$-wave pairing channel is vanishing due to $C_{4v}$ symmetry constraints. 
The model for the two-chains includes orbital dependent electronic charge transfer along the $x$ and $y$ as it is expected to occur for the ($xz$,$yz$) states in the presence of rotation and mirror symmetry breaking (see Supplemental Material \cite{SupMat} for the details of the model Hamiltonian). We introduce the amplitudes $t_{\gamma\gamma'}^{i}$ with $\gamma,\gamma'=a,b$ and $i=x,y$ to describe the effective hopping within the ($xz$,$yz$) manifold along the $x$ and $y$ directions. Due to the finite size of the model system, assuming that $t_{aa}=t_{bb}$ and $t_{ab}=0$, one recovers the rotational orbital symmetry and the density of states has the typical profile of a conventional $s$-wave superconductor (bottom panel in Fig.\ 4) with a full gap in the energy range $[-|\Delta_0|,|\Delta_0|]$. Then, one can break the orbital rotational symmetry by unbalancing the $t_{aa}$ and $t_{bb}$ amplitudes. With this assumption the $C_4$ rotation symmetry is broken while the vertical mirror symmetry is preserved. For instance, one can consider the change of the hopping amplitude of the $xz$ orbitals along the $y$ directions. The hopping asymmetry acts like an effective $k$-dependent orbital field in the ($xz$,$yz$) manifold and thus it splits the peaks associated with the orbital excitations. The splitting and the occurrence of spectral weight asymmetry can be observed in the density of states [Fig.\ 4(b)] already for a small deviation of $t^{y}_{bb}$ from $t^{y}_{aa}$. The spectral weight profile gets more asymmetric as the amplitude of the orbital field increases by unbalancing $t_{aa}$ with respect to $t_{bb}$. Similar modifications of the density of states can be obtained by varying the hopping amplitudes along the $x$ direction. 
Along this line, by considering a nonvanishing $t_{ab}$ one can also introduce a term that breaks both rotation and vertical mirror symmetries. As for the previous cases, such symmetry breaking yields effects on the density of states that are similar to those reported in Fig.\ 4. Additionally, one can also consider orbital triplet configurations in the orbital manifold and obtain particle-hole asymmetric spectra  in the presence of rotation and mirror symmetry breaking. 
\\
The emerging physical scenario can be also qualitatively accounted by making the following analogy. Assuming that the ($xz$,$yz$) states are configurations associated to a pseudospin $1/2$, the achieved spin-triplet orbital-singlet configuration in the presence of rotation and mirror symmetry breaking at the edge is analogous to a spin-singlet configuration -- spin up and down corresponding to the $xz$ and $yz$ orbitals -- subjected to a source of time reversal symmetry breaking with spin dependent mass in momentum space and anisotropic magnetic field. Thus, taking into account this analogy, the crystalline interactions that break rotation and mirror symmetries act as effective fields that break the time reversal in the pseudospin orbital space.

{\textit{Conclusions}}. 
We show the occurrence of asymmetric density of states in time-reversal spin-triplet superconductors equipped with an orbital degree of freedom that at the edge are subjected to a symmetry breaking field in the orbital space due to the reduced crystalline symmetry and orbital Andreev reflections.
The achieved outcomes have several consequences concerning the materials and the detection of unconventional pairing. 
First, the asymmetry in the DOS can be exploited to distinguish between acentric superconductors with time-reversal spin-triplet pairs in a single-orbital odd-parity channel (e.g.\ $p$-wave) with respect to those marked by a multi-orbital even-parity (e.g.\ $s$-wave) structure. Second, we have a clear-cut example of a superconductor equipped with Andreev bound states that results in asymmetric peaks in the spectra that depend on the strength of interactions that break the rotation and mirror symmetry. 
Such outcomes are in principle applicable to the recently discovered anomalies at the surface of the UTe$_2$ superconductor. While the occurrence of spin-triplet pairs in UTe$_2$ has been experimentally tested through nuclear magnetic resonance probes, our results indicate that the orbital degrees of freedom might play a role in setting out the structure of the spin-triplet pairing.
This is supported by ARPES observations demonstrating that in the normal state, apart from heavy $f$-bands, U 5$d$ and Te $p$ orbitals contribute to the electronic structure at the Fermi level \cite{Lin2020}.

Another application points to the class of 2D noncentrosymmetric oxides heterostructures, such as LaAlO$_3$/SrTiO$_3$ (LAO-STO)~\cite{Ohtomo-2004, Reyren-2007}, especially due to its remarkable orbital control of the superconducting critical temperature by electrostatic gating ~\cite{Caviglia-2008,Thierschmann-2018,Hurand-2015}, together with Rashba spin-orbit coupling~\cite{Caviglia2010PRL,Ben_Shalom2010PRL}, the occupation of the Ti 3d orbitals ($d_{xy}$,$d_{zx}$,$d_{yz}$)~\cite{Joshua-2012, Herranz-2015}, and the spin and orbital sources of Berry curvature~\cite{Lesne2023,Mercaldo2023}. 
The superconducting phase in LAO-STO exhibits several anomalous properties~\cite{Singh-2019,Trevisan-2018,Bal-2015,Stornaiuolo2017,Kalaboukhov-2017,Kuerten-2017,Sin21} that cannot be easily addressed within a conventional single orbital spin-singlet scenario.
Our findings indicate that the presence of an even-parity orbital dependent spin-triplet pairing would lead to gate tunable asymmetric spectral lineshape at the edges or interface in STO based heterostructures. This is expected in the regime of gating across the Lifshitz transition when the $(zx,yz)$ orbital manifold gets populated. 
We notice that similar effects are also expected at the surface of doped topological superconductors \cite{Mizushima2014} assuming two-orbital odd-parity spin-triplet pairing \cite{Fu_Berg_2010}. Moreover, we argue that our findings can be generalized to superconductors involving pseudospin degrees of freedom akin to orbitals (e.g.\ sublattice, valley) for terminations that yield a splitting of the pseudospin degrees of freedom. This can be achieved, for instance, in bilayer Rashba superconductors \cite{Nakosai2012}, or in transition metal dichalcogenides \cite{Xu2014}.

Finally, we would like to discuss possible experimental proposals in order to single out the asymmetric features of the density of states arising from the inter-orbital spin-triplet pairing. In this context, the first aspect to consider is that our findings indicate that the asymmetric profile arises in the presence of time-reversal symmetry. Hence, from an experimental perspective, in order to isolate the uncovered orbital dependent features, one has to avoid magnetic fields and preliminarily test that the normal state is not magnetic. In principle, magnetic impurities can lead to Kondo resonances close to the Fermi level which are marked by an asymmetric profile in the density of states. However, the energy scales of the Kondo features are typically larger than the superconducting gap \cite{JiaoNature2020} and their evolution in the presence of an applied magnetic field can help to discern them from those arising from the superconducting excitations.
Instead, since the orbital degrees and the reduced crystalline symmetry play an important role in steering the asymmetric profile of the spectral function for multiorbital superconductors, we argue that the application of strain \cite{Mutch2019} or electric fields \cite{Simoni2018, Simoni2021, Mercaldo2021} can be valuable means to investigate the occurrence of particle-hole asymmetry in the density of states. This is because both strain and gate fields do not break the time-reversal symmetry and one can implement suitable setups to control the strength of the dissipation effects. Having a profile of the density of states at the edge whose asymmetry is augmented or reduced by applying strain or gate fields may be an indirect evidence of orbital dependent spin-triplet pairing.
\\
Another path to singling out the pairing induced particle-hole asymmetry is to exploit the tunability of the orbital population at the Fermi level. For instance, in multiorbital superconductors as LAO-STO heterostructures, the application of gate voltage can tune the orbital occupation from a dominant $xy$ character to a combination of $xy$ and $xz,yz$ orbital flavor. In this case, the occurrence of an asymmetric density of states at the edge in concomitance with the changeover of the orbital population can be used to single out the orbital dependent spin-triplet pairing.
\\
Finally, the potential to induce gate driven asymmetry in the density of states can be exploited in junctions based on orbital dependent spin-triplet pairing to achieve nonstandard magneto-electric and topological effects, as compared to the case of single band p-wave spin-triplet superconductors \cite{MKC2019,MKC2018}, as well as to uncoventional orbital Edelstein effects \cite{Chirolli}.


\textit{Acknowledgements}.\ 
M.C., P.G.\ and Y.F.\ acknowledge support by the project ``Two-dimensional Oxides Platform for SPINorbitronics nanotechnology (TOPSPIN)'' funded by the MIUR-PRIN Bando 2017 - Grant No.\ 20177SL7HC.
M.C.\ and Y.F.\ acknowledge support by the EU’s Horizon 2020 research and innovation program under Grant Agreement No.\ 964398 (SUPERGATE).
This work was supported by JSPS KAKENHI Grants No.\ JP20H00131, No.\ JP18H01176, and No.\ JP20H01857, and the JSPS-EPSRC Core-to-Core program ``Oxide Superspin'' international network (Grant No.\ JPJSCCA20170002).



\begin{thebibliography}{66}%
\makeatletter
\providecommand \@ifxundefined [1]{%
 \@ifx{#1\undefined}
}%
\providecommand \@ifnum [1]{%
 \ifnum #1\expandafter \@firstoftwo
 \else \expandafter \@secondoftwo
 \fi
}%
\providecommand \@ifx [1]{%
 \ifx #1\expandafter \@firstoftwo
 \else \expandafter \@secondoftwo
 \fi
}%
\providecommand \natexlab [1]{#1}%
\providecommand \enquote  [1]{``#1''}%
\providecommand \bibnamefont  [1]{#1}%
\providecommand \bibfnamefont [1]{#1}%
\providecommand \citenamefont [1]{#1}%
\providecommand \href@noop [0]{\@secondoftwo}%
\providecommand \href [0]{\begingroup \@sanitize@url \@href}%
\providecommand \@href[1]{\@@startlink{#1}\@@href}%
\providecommand \@@href[1]{\endgroup#1\@@endlink}%
\providecommand \@sanitize@url [0]{\catcode `\\12\catcode `\$12\catcode
  `\&12\catcode `\#12\catcode `\^12\catcode `\_12\catcode `\%12\relax}%
\providecommand \@@startlink[1]{}%
\providecommand \@@endlink[0]{}%
\providecommand \url  [0]{\begingroup\@sanitize@url \@url }%
\providecommand \@url [1]{\endgroup\@href {#1}{\urlprefix }}%
\providecommand \urlprefix  [0]{URL }%
\providecommand \Eprint [0]{\href }%
\providecommand \doibase [0]{https://doi.org/}%
\providecommand \selectlanguage [0]{\@gobble}%
\providecommand \bibinfo  [0]{\@secondoftwo}%
\providecommand \bibfield  [0]{\@secondoftwo}%
\providecommand \translation [1]{[#1]}%
\providecommand \BibitemOpen [0]{}%
\providecommand \bibitemStop [0]{}%
\providecommand \bibitemNoStop [0]{.\EOS\space}%
\providecommand \EOS [0]{\spacefactor3000\relax}%
\providecommand \BibitemShut  [1]{\csname bibitem#1\endcsname}%
\let\auto@bib@innerbib\@empty
\bibitem [{\citenamefont {Andreev}(1964)}]{Andreev_reflection}%
  \BibitemOpen
  \bibfield  {author} {\bibinfo {author} {\bibfnamefont {A.}~\bibnamefont
  {Andreev}},\ }\bibfield  {title} {\bibinfo {title} {The thermal conductivity
  of the intermediate state in superconductors},\ }\href
  {https://doi.org/10.1098/rsta.2018.0140} {\bibfield  {journal}
  {\bibinfo  {journal} {Sov. Phys. JETP}\ }\textbf {\bibinfo {volume} {19}},\
  \bibinfo {pages} {1228} (\bibinfo {year} {1964})}\BibitemShut {NoStop}%
\bibitem [{\citenamefont {Bruder}(1990)}]{Bruder1990}%
  \BibitemOpen
  \bibfield  {author} {\bibinfo {author} {\bibfnamefont {C.}~\bibnamefont
  {Bruder}},\ }\bibfield  {title} {\bibinfo {title} {Andreev scattering in
  anisotropic superconductors},\ }\href
  {https://doi.org/10.1103/PhysRevB.41.4017} {\bibfield  {journal} {\bibinfo
  {journal} {Phys. Rev. B}\ }\textbf {\bibinfo {volume} {41}},\ \bibinfo
  {pages} {4017} (\bibinfo {year} {1990})}\BibitemShut {NoStop}%
\bibitem [{\citenamefont {Kashiwaya}\ and\ \citenamefont
  {Tanaka}(2000)}]{KashiwayaTanaka2000RepProgPhys}%
  \BibitemOpen
  \bibfield  {author} {\bibinfo {author} {\bibfnamefont {S.}~\bibnamefont
  {Kashiwaya}}\ and\ \bibinfo {author} {\bibfnamefont {Y.}~\bibnamefont
  {Tanaka}},\ }\bibfield  {title} {\bibinfo {title} {Tunnelling effects on
  surface bound states in unconventional superconductors},\ }\href@noop {}
  {\bibfield  {journal} {\bibinfo  {journal} {Rep. Prog. Phys.}\ }\textbf
  {\bibinfo {volume} {63}},\ \bibinfo {pages} {1641} (\bibinfo {year}
  {2000})}\BibitemShut {NoStop}%
\bibitem [{\citenamefont {Sauls}(2018)}]{Sauls2018}%
  \BibitemOpen
  \bibfield  {author} {\bibinfo {author} {\bibfnamefont {J.~A.}\ \bibnamefont
  {Sauls}},\ }\bibfield  {title} {\bibinfo {title} {Andreev bound states and
  their signatures},\ }\href@noop {} {\bibfield  {journal} {\bibinfo  {journal}
  {Philosophical Transactions of the Royal Society of London A}\ }\textbf
  {\bibinfo {volume} {376}},\ \bibinfo {pages} {2125} (\bibinfo {year}
  {2018})}\BibitemShut {NoStop}%
\bibitem [{\citenamefont {Prada}\ \emph {et~al.}(2020)\citenamefont {Prada},
  \citenamefont {San-Jose}, \citenamefont {de~Moor}, \citenamefont {Geresdi},
  \citenamefont {Lee}, \citenamefont {Klinovaja}, \citenamefont {Loss},
  \citenamefont {Nygard}, \citenamefont {Aguado},\ and\ \citenamefont
  {Kouwenhoven}}]{Prada2020}%
  \BibitemOpen
  \bibfield  {author} {\bibinfo {author} {\bibfnamefont {E.}~\bibnamefont
  {Prada}}, \bibinfo {author} {\bibfnamefont {P.}~\bibnamefont {San-Jose}},
  \bibinfo {author} {\bibfnamefont {M.~W.~A.}\ \bibnamefont {de~Moor}},
  \bibinfo {author} {\bibfnamefont {A.}~\bibnamefont {Geresdi}}, \bibinfo
  {author} {\bibfnamefont {E.~J.~H.}\ \bibnamefont {Lee}}, \bibinfo {author}
  {\bibfnamefont {J.}~\bibnamefont {Klinovaja}}, \bibinfo {author}
  {\bibfnamefont {D.}~\bibnamefont {Loss}}, \bibinfo {author} {\bibfnamefont
  {J.}~\bibnamefont {Nygard}}, \bibinfo {author} {\bibfnamefont
  {R.}~\bibnamefont {Aguado}},\ and\ \bibinfo {author} {\bibfnamefont {L.~P.}\
  \bibnamefont {Kouwenhoven}},\ }\bibfield  {title} {\bibinfo {title} {From
  Andreev to Majorana bound states in hybrid superconductor–semiconductor
  nanowires},\ }\href@noop {} {\bibfield  {journal} {\bibinfo  {journal} {Nat.
  Rev. Phys.}\ }\textbf {\bibinfo {volume} {2}},\ \bibinfo {pages} {575}
  (\bibinfo {year} {2020})}\BibitemShut {NoStop}%
\bibitem [{\citenamefont {Tanaka}\ (1995)\citenamefont {Tanaka}\ and\ \citenamefont {Kashiwaya}}]{tanaka1995PRL}%
  \BibitemOpen
  \bibfield  {author} {\bibinfo {author} {\bibfnamefont {Y.}~\bibnamefont
  {Tanaka}}\ and\
  \bibinfo {author} {\bibfnamefont {S.}~\bibnamefont {kashiwaya}},\ }\bibfield
  {title} {\bibinfo {title} {Theory of Tunneling Spectroscopy of d-Wave Superconductors},\ }\href
  {https://doi.org/10.1103/PhysRevLett.74.3451} {\bibfield  {journal} {\bibinfo
  {journal} {Phys. Rev. Lett.}\ }\textbf {\bibinfo {volume} {74}},\ \bibinfo
  {pages} {3451} (\bibinfo {year} {1995})}\BibitemShut {NoStop}%
\bibitem [{\citenamefont {Qi}\ and\ \citenamefont {Zhang}(2011)}]{Qi2011}%
  \BibitemOpen
  \bibfield  {author} {\bibinfo {author} {\bibfnamefont {X.-L.}\ \bibnamefont
  {Qi}}\ and\ \bibinfo {author} {\bibfnamefont {S.-C.}\ \bibnamefont {Zhang}},\
  }\bibfield  {title} {\bibinfo {title} {Topological insulators and
  superconductors},\ }\href@noop {} {\bibfield  {journal} {\bibinfo  {journal}
  {Rev. Mod. Phys.}\ }\textbf {\bibinfo {volume} {83}},\ \bibinfo {pages}
  {1057} (\bibinfo {year} {2011})}\BibitemShut {NoStop}%
\bibitem [{\citenamefont {Sato}\ and\ \citenamefont {Ando}(2017)}]{sato2017}%
  \BibitemOpen
  \bibfield  {author} {\bibinfo {author} {\bibfnamefont {M.}~\bibnamefont
  {Sato}}\ and\ \bibinfo {author} {\bibfnamefont {Y.}~\bibnamefont {Ando}},\
  }\bibfield  {title} {\bibinfo {title} {Topological superconductors: a
  review},\ }\href@noop {} {\bibfield  {journal} {\bibinfo  {journal} {Rep.
  Prog. Phys.}\ }\textbf {\bibinfo {volume} {80}},\ \bibinfo {pages} {076501}
  (\bibinfo {year} {2017})}\BibitemShut {NoStop}%
\bibitem [{\citenamefont {Flensberg}\ \emph {et~al.}(2021)\citenamefont
  {Flensberg}, \citenamefont {von Oppen},\ and\ \citenamefont
  {Stern}}]{Flensberg2021}%
  \BibitemOpen
  \bibfield  {author} {\bibinfo {author} {\bibfnamefont {K.}~\bibnamefont
  {Flensberg}}, \bibinfo {author} {\bibfnamefont {F.}~\bibnamefont {von
  Oppen}},\ and\ \bibinfo {author} {\bibfnamefont {A.}~\bibnamefont {Stern}},\
  }\bibfield  {title} {\bibinfo {title} {Engineered platforms for topological
  superconductivity and Majorana zero modes},\ }\href@noop {} {\bibfield
  {journal} {\bibinfo  {journal} {Nat. Rev. Mat.}\ }\textbf {\bibinfo {volume}
  {6}},\ \bibinfo {pages} {944} (\bibinfo {year} {2021})}\BibitemShut {NoStop}%
\bibitem [{\citenamefont {Nayak}\ \emph {et~al.}(2008)\citenamefont {Nayak},
  \citenamefont {Simon}, \citenamefont {Stern}, \citenamefont {Freedman},\ and\
  \citenamefont {Das~Sarma}}]{Nayak2008}%
  \BibitemOpen
  \bibfield  {author} {\bibinfo {author} {\bibfnamefont {C.}~\bibnamefont
  {Nayak}}, \bibinfo {author} {\bibfnamefont {S.~H.}\ \bibnamefont {Simon}},
  \bibinfo {author} {\bibfnamefont {A.}~\bibnamefont {Stern}}, \bibinfo
  {author} {\bibfnamefont {M.}~\bibnamefont {Freedman}},\ and\ \bibinfo
  {author} {\bibfnamefont {S.}~\bibnamefont {Das~Sarma}},\ }\bibfield  {title}
  {\bibinfo {title} {Non-abelian anyons and topological quantum computation},\
  }\href@noop {} {\bibfield  {journal} {\bibinfo  {journal} {Rev. Mod. Phys.}\
  }\textbf {\bibinfo {volume} {80}},\ \bibinfo {pages} {1083} (\bibinfo {year}
  {2008})}\BibitemShut {NoStop}%
\bibitem [{\citenamefont {Ran}\ \emph {et~al.}(2019)\citenamefont {Ran},
  \citenamefont {Eckberg}, \citenamefont {Ding}, \citenamefont {Furukawa},
  \citenamefont {Metz}, \citenamefont {Saha}, \citenamefont {Liu},
  \citenamefont {Zic}, \citenamefont {Kim}, \citenamefont {Paglione},\ and\
  \citenamefont {Butch}}]{RanScience2019}%
  \BibitemOpen
  \bibfield  {author} {\bibinfo {author} {\bibfnamefont {S.}~\bibnamefont
  {Ran}}, \bibinfo {author} {\bibfnamefont {C.}~\bibnamefont {Eckberg}},
  \bibinfo {author} {\bibfnamefont {Q.-P.}\ \bibnamefont {Ding}}, \bibinfo
  {author} {\bibfnamefont {Y.}~\bibnamefont {Furukawa}}, \bibinfo {author}
  {\bibfnamefont {T.}~\bibnamefont {Metz}}, \bibinfo {author} {\bibfnamefont
  {S.~R.}\ \bibnamefont {Saha}}, \bibinfo {author} {\bibfnamefont {I.-L.}\
  \bibnamefont {Liu}}, \bibinfo {author} {\bibfnamefont {M.}~\bibnamefont
  {Zic}}, \bibinfo {author} {\bibfnamefont {H.}~\bibnamefont {Kim}}, \bibinfo
  {author} {\bibfnamefont {J.}~\bibnamefont {Paglione}},\ and\ \bibinfo
  {author} {\bibfnamefont {N.~P.}\ \bibnamefont {Butch}},\ }\bibfield  {title}
  {\bibinfo {title} {Nearly ferromagnetic spin-triplet superconductivity},\
  }\href {https://doi.org/10.1126/science.aav8645} {\bibfield  {journal}
  {\bibinfo  {journal} {Science}\ }\textbf {\bibinfo {volume} {365}},\ \bibinfo
  {pages} {684} (\bibinfo {year} {2019})}\BibitemShut {NoStop}%
\bibitem [{\citenamefont {Aoki}\ \emph {et~al.}(2019)\citenamefont {Aoki},
  \citenamefont {Nakamura}, \citenamefont {Honda}, \citenamefont {Li},
  \citenamefont {Homma}, \citenamefont {Shimizu}, \citenamefont {Sato},
  \citenamefont {Knebel}, \citenamefont {Brison}, \citenamefont {Pourret},
  \citenamefont {Braithwaite}, \citenamefont {Lapertot}, \citenamefont {Niu},
  \citenamefont {Valiska}, \citenamefont {Harima},\ and\ \citenamefont
  {Flouquet}}]{DAokiJPSJ2019}%
  \BibitemOpen
  \bibfield  {author} {\bibinfo {author} {\bibfnamefont {D.}~\bibnamefont
  {Aoki}}, \bibinfo {author} {\bibfnamefont {A.}~\bibnamefont {Nakamura}},
  \bibinfo {author} {\bibfnamefont {F.}~\bibnamefont {Honda}}, \bibinfo
  {author} {\bibfnamefont {D.}~\bibnamefont {Li}}, \bibinfo {author}
  {\bibfnamefont {Y.}~\bibnamefont {Homma}}, \bibinfo {author} {\bibfnamefont
  {Y.}~\bibnamefont {Shimizu}}, \bibinfo {author} {\bibfnamefont {a.~J.}\
  \bibnamefont {Sato}}, \bibinfo {author} {\bibfnamefont {G.}~\bibnamefont
  {Knebel}}, \bibinfo {author} {\bibfnamefont {J.-P.}\ \bibnamefont {Brison}},
  \bibinfo {author} {\bibfnamefont {A.}~\bibnamefont {Pourret}}, \bibinfo
  {author} {\bibfnamefont {D.}~\bibnamefont {Braithwaite}}, \bibinfo {author}
  {\bibfnamefont {G.}~\bibnamefont {Lapertot}}, \bibinfo {author}
  {\bibfnamefont {Q.}~\bibnamefont {Niu}}, \bibinfo {author} {\bibfnamefont
  {M.}~\bibnamefont {Valiska}}, \bibinfo {author} {\bibfnamefont
  {H.}~\bibnamefont {Harima}},\ and\ \bibinfo {author} {\bibfnamefont
  {J.}~\bibnamefont {Flouquet}},\ }\bibfield  {title} {\bibinfo {title}
  {Unconventional superconductivity in heavy fermion UTe$_2$},\ }\href
  {https://doi.org/10.7566/JPSJ.88.043702} {\bibfield  {journal} {\bibinfo
  {journal} {J. Phys. Soc. Jpn.}\ }\textbf {\bibinfo {volume} {88}},\ \bibinfo
  {pages} {043702} (\bibinfo {year} {2019})}\BibitemShut {NoStop}%
\bibitem [{\citenamefont {Aoki}\ \emph {et~al.}(2022)\citenamefont {Aoki},
  \citenamefont {Brison}, \citenamefont {Flouquet}, \citenamefont {Ishida},
  \citenamefont {Tokunaga},\ and\ \citenamefont
  {Yanase}}]{DAokiJofP:CondMatter2022}%
  \BibitemOpen
  \bibfield  {author} {\bibinfo {author} {\bibfnamefont {D.}~\bibnamefont
  {Aoki}}, \bibinfo {author} {\bibfnamefont {J.-P.}\ \bibnamefont {Brison}},
  \bibinfo {author} {\bibfnamefont {J.}~\bibnamefont {Flouquet}}, \bibinfo
  {author} {\bibfnamefont {K.}~\bibnamefont {Ishida}}, \bibinfo {author}
  {\bibfnamefont {Y.}~\bibnamefont {Tokunaga}},\ and\ \bibinfo {author}
  {\bibfnamefont {Y.}~\bibnamefont {Yanase}},\ }\bibfield  {title} {\bibinfo
  {title} {Unconventional superconductivity in UTe$_2$},\ }\href
  {https://doi.org/10.1088/1361-648X/ac5863} {\bibfield  {journal} {\bibinfo
  {journal} {J. Phys.: Condens. Matter}\ }\textbf {\bibinfo {volume} {34}},\
  \bibinfo {pages} {243002} (\bibinfo {year} {2022})}\BibitemShut {NoStop}%
\bibitem [{\citenamefont {Jiao}\ \emph {et~al.}(2020)\citenamefont {Jiao},
  \citenamefont {Howard}, \citenamefont {Ran}, \citenamefont {Wang},
  \citenamefont {Rodriguez}, \citenamefont {Sigrist}, \citenamefont {Wang},
  \citenamefont {Butch},\ and\ \citenamefont {Madhavan}}]{JiaoNature2020}%
  \BibitemOpen
  \bibfield  {author} {\bibinfo {author} {\bibfnamefont {L.}~\bibnamefont
  {Jiao}}, \bibinfo {author} {\bibfnamefont {S.}~\bibnamefont {Howard}},
  \bibinfo {author} {\bibfnamefont {S.}~\bibnamefont {Ran}}, \bibinfo {author}
  {\bibfnamefont {Z.}~\bibnamefont {Wang}}, \bibinfo {author} {\bibfnamefont
  {J.~O.}\ \bibnamefont {Rodriguez}}, \bibinfo {author} {\bibfnamefont
  {M.}~\bibnamefont {Sigrist}}, \bibinfo {author} {\bibfnamefont
  {Z.}~\bibnamefont {Wang}}, \bibinfo {author} {\bibfnamefont {N.~P.}\
  \bibnamefont {Butch}},\ and\ \bibinfo {author} {\bibfnamefont
  {V.}~\bibnamefont {Madhavan}},\ }\bibfield  {title} {\bibinfo {title} {Chiral
  superconductivity in heavy-fermion metal UTe$_2$},\ }\href
  {https://doi.org/10.1038/s41586-020-2122-2} {\bibfield  {journal} {\bibinfo
  {journal} {Nature}\ }\textbf {\bibinfo {volume} {579}},\ \bibinfo {pages}
  {523} (\bibinfo {year} {2020})}\BibitemShut {NoStop}%
\bibitem [{\citenamefont {Tanuma}\ \emph {et~al.}(1998)\citenamefont {Tanuma},
  \citenamefont {Tanaka}, \citenamefont {Yamashiro},\ and\ \citenamefont
  {Kashiwaya}}]{Tanuma1997}%
  \BibitemOpen
  \bibfield  {author} {\bibinfo {author} {\bibfnamefont {Y.}~\bibnamefont
  {Tanuma}}, \bibinfo {author} {\bibfnamefont {Y.}~\bibnamefont {Tanaka}},
  \bibinfo {author} {\bibfnamefont {M.}~\bibnamefont {Yamashiro}},\ and\
  \bibinfo {author} {\bibfnamefont {S.}~\bibnamefont {Kashiwaya}},\ }\bibfield
  {title} {\bibinfo {title} {Local density of states on rough surfaces of
  ${d}_{{x}^{2}\ensuremath{-}{y}^{2}}$-wave superconductors},\ }\href
  {https://doi.org/10.1103/PhysRevB.57.7997} {\bibfield  {journal} {\bibinfo
  {journal} {Phys. Rev. B}\ }\textbf {\bibinfo {volume} {57}},\ \bibinfo
  {pages} {7997} (\bibinfo {year} {1998})}\BibitemShut {NoStop}%
\bibitem [{\citenamefont {Tanuma}\ \emph {et~al.}(1999)\citenamefont {Tanuma},
  \citenamefont {Tanaka}, \citenamefont {Ogata},\ and\ \citenamefont
  {Kashiwaya}}]{Tanuma1999}%
  \BibitemOpen
  \bibfield  {author} {\bibinfo {author} {\bibfnamefont {Y.}~\bibnamefont
  {Tanuma}}, \bibinfo {author} {\bibfnamefont {Y.}~\bibnamefont {Tanaka}},
  \bibinfo {author} {\bibfnamefont {M.}~\bibnamefont {Ogata}},\ and\ \bibinfo
  {author} {\bibfnamefont {S.}~\bibnamefont {Kashiwaya}},\ }\bibfield  {title}
  {\bibinfo {title} {Quasiparticle states near surfaces of
  $\mathrm{high}{\ensuremath{-}T}_{c}$ superconductors based on the extended
  $t\ensuremath{-}j$ model},\ }\href {https://doi.org/10.1103/PhysRevB.60.9817}
  {\bibfield  {journal} {\bibinfo  {journal} {Phys. Rev. B}\ }\textbf {\bibinfo
  {volume} {60}},\ \bibinfo {pages} {9817} (\bibinfo {year}
  {1999})}\BibitemShut {NoStop}%
\bibitem [{\citenamefont {Yang}\ \emph {et~al.}(2007)\citenamefont {Yang},
  \citenamefont {Yang}, \citenamefont {Jiang},\ and\ \citenamefont
  {Li}}]{YangJofP:cond_matter2007}%
  \BibitemOpen
  \bibfield  {author} {\bibinfo {author} {\bibfnamefont {H.-Y.}\ \bibnamefont
  {Yang}}, \bibinfo {author} {\bibfnamefont {F.}~\bibnamefont {Yang}}, \bibinfo
  {author} {\bibfnamefont {Y.-J.}\ \bibnamefont {Jiang}},\ and\ \bibinfo
  {author} {\bibfnamefont {T.}~\bibnamefont {Li}},\ }\bibfield  {title}
  {\bibinfo {title} {On the origin of the tunnelling asymmetry in the cuprate
  superconductors: a variational perspective},\ }\href
  {https://doi.org/10.1088/0953-8984/19/1/016217} {\bibfield  {journal}
  {\bibinfo  {journal} {J. Phys: Condens. Matter}\ }\textbf {\bibinfo {volume}
  {19}},\ \bibinfo {pages} {016217} (\bibinfo {year} {2007})}\BibitemShut
  {NoStop}%
\bibitem [{\citenamefont {Zou}\ \emph {et~al.}(2022)\citenamefont {Zou},
  \citenamefont {Hao}, \citenamefont {Luo}, \citenamefont {Ye}, \citenamefont
  {Gao}, \citenamefont {Li}, \citenamefont {Xu}, \citenamefont {Cai},
  \citenamefont {Lin}, \citenamefont {Zhou}, \citenamefont {Lee},\ and\
  \citenamefont {Wang}}]{Zou2021arXiv}%
  \BibitemOpen
  \bibfield  {author} {\bibinfo {author} {\bibfnamefont {C.}~\bibnamefont
  {Zou}}, \bibinfo {author} {\bibfnamefont {Z.}~\bibnamefont {Hao}}, \bibinfo
  {author} {\bibfnamefont {X.}~\bibnamefont {Luo}}, \bibinfo {author}
  {\bibfnamefont {S.}~\bibnamefont {Ye}}, \bibinfo {author} {\bibfnamefont
  {Q.}~\bibnamefont {Gao}}, \bibinfo {author} {\bibfnamefont {X.}~\bibnamefont
  {Li}}, \bibinfo {author} {\bibfnamefont {M.}~\bibnamefont {Xu}}, \bibinfo
  {author} {\bibfnamefont {P.}~\bibnamefont {Cai}}, \bibinfo {author}
  {\bibfnamefont {C.}~\bibnamefont {Lin}}, \bibinfo {author} {\bibfnamefont
  {X.}~\bibnamefont {Zhou}}, \bibinfo {author} {\bibfnamefont {D.-H.}\
  \bibnamefont {Lee}},\ and\ \bibinfo {author} {\bibfnamefont {Y.}~\bibnamefont
  {Wang}},\ }\href@noop {} {\bibinfo {title} {Particle-hole asymmetric
  superconducting coherence peaks in overdoped cuprates}},\ \href{https://doi.org/10.1038/s41567-022-01534-x} {\bibfield  {journal} {\bibinfo
  {journal} {Nat. Phys.}\ }\textbf {\bibinfo {volume} {18}},\ \bibinfo
  {pages} {551} (\bibinfo {year} {2022})} \BibitemShut {NoStop}%
\bibitem [{\citenamefont {Ng}\ and\ \citenamefont {Varma}(2004)}]{NgPRB2004}%
  \BibitemOpen
  \bibfield  {author} {\bibinfo {author} {\bibfnamefont {T.-K.}\ \bibnamefont
  {Ng}}\ and\ \bibinfo {author} {\bibfnamefont {C.~M.}\ \bibnamefont {Varma}},\
  }\bibfield  {title} {\bibinfo {title} {Experimental signatures of
  time-reversal-violating superconductors},\ }\href
  {https://doi.org/10.1103/PhysRevB.70.054514} {\bibfield  {journal} {\bibinfo
  {journal} {Phys. Rev. B}\ }\textbf {\bibinfo {volume} {70}},\ \bibinfo
  {pages} {054514} (\bibinfo {year} {2004})}\BibitemShut {NoStop}%
\bibitem [{\citenamefont {Linder}\ \emph {et~al.}(2010)\citenamefont {Linder},
  \citenamefont {Cuoco},\ and\ \citenamefont {Sudb\o{}}}]{Linder2010}%
  \BibitemOpen
  \bibfield  {author} {\bibinfo {author} {\bibfnamefont {J.}~\bibnamefont
  {Linder}}, \bibinfo {author} {\bibfnamefont {M.}~\bibnamefont {Cuoco}},\ and\
  \bibinfo {author} {\bibfnamefont {A.}~\bibnamefont {Sudb\o{}}},\ }\bibfield
  {title} {\bibinfo {title} {Spin-active interfaces and unconventional pairing
  in half-metal/superconductor junctions},\ }\href
  {https://doi.org/10.1103/PhysRevB.81.174526} {\bibfield  {journal} {\bibinfo
  {journal} {Phys. Rev. B}\ }\textbf {\bibinfo {volume} {81}},\ \bibinfo
  {pages} {174526} (\bibinfo {year} {2010})}\BibitemShut {NoStop}%
\bibitem [{\citenamefont {Eschrig}(2015)}]{EschrigRPP2015}%
  \BibitemOpen
  \bibfield  {author} {\bibinfo {author} {\bibfnamefont {M.}~\bibnamefont
  {Eschrig}},\ }\bibfield  {title} {\bibinfo {title} {Spin-polarized
  supercurrents for spintronics},\ }\href@noop {} {\bibfield  {journal}
  {\bibinfo  {journal} {Rep. Prog. Phys.}\ }\textbf {\bibinfo {volume} {78}},\
  \bibinfo {pages} {104501} (\bibinfo {year} {2015})}\BibitemShut {NoStop}%
\bibitem [{\citenamefont {Lu}\ \emph {et~al.}(2015)\citenamefont {Lu},
  \citenamefont {Burset}, \citenamefont {Yada},\ and\ \citenamefont
  {Tanaka}}]{Lu_2015}%
  \BibitemOpen
  \bibfield  {author} {\bibinfo {author} {\bibfnamefont {B.}~\bibnamefont
  {Lu}}, \bibinfo {author} {\bibfnamefont {P.}~\bibnamefont {Burset}}, \bibinfo
  {author} {\bibfnamefont {K.}~\bibnamefont {Yada}},\ and\ \bibinfo {author}
  {\bibfnamefont {Y.}~\bibnamefont {Tanaka}},\ }\bibfield  {title} {\bibinfo
  {title} {Tunneling spectroscopy and josephson current of
  superconductor-ferromagnet hybrids on the surface of a 3d Ti},\ }\href
  {https://doi.org/10.1088/0953-2048/28/10/105001} {\bibfield  {journal}
  {\bibinfo  {journal} {Superconductor Science and Technology}\ }\textbf
  {\bibinfo {volume} {28}},\ \bibinfo {pages} {105001} (\bibinfo {year}
  {2015})}\BibitemShut {NoStop}%
\bibitem [{\citenamefont {Lu}\ and\ \citenamefont {Tanaka}(2018)}]{Lu_2018}%
  \BibitemOpen
  \bibfield  {author} {\bibinfo {author} {\bibfnamefont {B.}~\bibnamefont
  {Lu}}\ and\ \bibinfo {author} {\bibfnamefont {Y.}~\bibnamefont {Tanaka}},\
  }\bibfield  {title} {\bibinfo {title} {Study on Green’s function on
  topological insulator surface},\ }\href@noop {} {\bibfield  {journal}
  {\bibinfo  {journal} {Phil. Trans. R. Soc. A}\ }\textbf {\bibinfo {volume}
  {376}},\ \bibinfo {pages} {20150246} (\bibinfo {year} {2018})}\BibitemShut
  {NoStop}%
\bibitem [{\citenamefont {Ando}\ \emph {et~al.}(2022)\citenamefont {Ando},
  \citenamefont {Ikegaya}, \citenamefont {Tamura}, \citenamefont {Tanaka},\
  and\ \citenamefont {Yada}}]{Ando2022}%
  \BibitemOpen
  \bibfield  {author} {\bibinfo {author} {\bibfnamefont {S.}~\bibnamefont
  {Ando}}, \bibinfo {author} {\bibfnamefont {S.}~\bibnamefont {Ikegaya}},
  \bibinfo {author} {\bibfnamefont {S.}~\bibnamefont {Tamura}}, \bibinfo
  {author} {\bibfnamefont {Y.}~\bibnamefont {Tanaka}},\ and\ \bibinfo {author}
  {\bibfnamefont {K.}~\bibnamefont {Yada}},\ }\bibfield  {title} {\bibinfo
  {title} {Surface state of the interorbital pairing state in the
  ${\mathrm{Sr}}_{2}{\mathrm{RuO}}_{4}$ superconductor},\ }\href
  {https://doi.org/10.1103/PhysRevB.106.214520} {\bibfield  {journal} {\bibinfo
   {journal} {Phys. Rev. B}\ }\textbf {\bibinfo {volume} {106}},\ \bibinfo
  {pages} {214520} (\bibinfo {year} {2022})}\BibitemShut {NoStop}%
\bibitem [{\citenamefont {Setiawan}\ and\ \citenamefont
  {Sau}(2021)}]{SetiawanPRR2021}%
  \BibitemOpen
  \bibfield  {author} {\bibinfo {author} {\bibfnamefont {F.}~\bibnamefont
  {Setiawan}}\ and\ \bibinfo {author} {\bibfnamefont {J.~D.}\ \bibnamefont
  {Sau}},\ }\bibfield  {title} {\bibinfo {title} {Electron-boson-interaction
  induced particle-hole symmetry breaking of conductance into subgap states in
  superconductors},\ }\href {https://doi.org/10.1103/PhysRevResearch.3.L032038}
  {\bibfield  {journal} {\bibinfo  {journal} {Phys. Rev. Research}\ }\textbf
  {\bibinfo {volume} {3}},\ \bibinfo {pages} {L032038} (\bibinfo {year}
  {2021})}\BibitemShut {NoStop}%
\bibitem [{\citenamefont {Cuoco}\ and\ \citenamefont
  {Ranninger}(2004)}]{Cuoco2004}%
  \BibitemOpen
  \bibfield  {author} {\bibinfo {author} {\bibfnamefont {M.}~\bibnamefont
  {Cuoco}}\ and\ \bibinfo {author} {\bibfnamefont {J.}~\bibnamefont
  {Ranninger}},\ }\bibfield  {title} {\bibinfo {title}
  {Superconductor-insulator transition driven by local dephasing},\ }\href
  {https://doi.org/10.1103/PhysRevB.70.104509} {\bibfield  {journal} {\bibinfo
  {journal} {Phys. Rev. B}\ }\textbf {\bibinfo {volume} {70}},\ \bibinfo
  {pages} {104509} (\bibinfo {year} {2004})}\BibitemShut {NoStop}%
\bibitem [{\citenamefont {Cuoco}\ and\ \citenamefont
  {Ranninger}(2006)}]{Cuoco2006}%
  \BibitemOpen
  \bibfield  {author} {\bibinfo {author} {\bibfnamefont {M.}~\bibnamefont
  {Cuoco}}\ and\ \bibinfo {author} {\bibfnamefont {J.}~\bibnamefont
  {Ranninger}},\ }\bibfield  {title} {\bibinfo {title} {From an insulating to a
  superfluid pair-bond liquid},\ }\href
  {https://doi.org/10.1103/PhysRevB.74.094511} {\bibfield  {journal} {\bibinfo
  {journal} {Phys. Rev. B}\ }\textbf {\bibinfo {volume} {74}},\ \bibinfo
  {pages} {094511} (\bibinfo {year} {2006})}\BibitemShut {NoStop}%
\bibitem [{\citenamefont {Woods}\ \emph {et~al.}(2019)\citenamefont {Woods},
  \citenamefont {Chen}, \citenamefont {Frolov},\ and\ \citenamefont
  {Stanescu}}]{Woods2019}%
  \BibitemOpen
  \bibfield  {author} {\bibinfo {author} {\bibfnamefont {B.~D.}\ \bibnamefont
  {Woods}}, \bibinfo {author} {\bibfnamefont {J.}~\bibnamefont {Chen}},
  \bibinfo {author} {\bibfnamefont {S.~M.}\ \bibnamefont {Frolov}},\ and\
  \bibinfo {author} {\bibfnamefont {T.~D.}\ \bibnamefont {Stanescu}},\
  }\bibfield  {title} {\bibinfo {title} {Zero-energy pinning of topologically
  trivial bound states in multiband semiconductor-superconductor nanowires},\
  }\href {https://doi.org/10.1103/PhysRevB.100.125407} {\bibfield  {journal}
  {\bibinfo  {journal} {Phys. Rev. B}\ }\textbf {\bibinfo {volume} {100}},\
  \bibinfo {pages} {125407} (\bibinfo {year} {2019})}\BibitemShut {NoStop}%
\bibitem [{\citenamefont {Liu}\ \emph {et~al.}(2017)\citenamefont {Liu},
  \citenamefont {Sau},\ and\ \citenamefont {Das~Sarma}}]{Liu2017}%
  \BibitemOpen
  \bibfield  {author} {\bibinfo {author} {\bibfnamefont {C.-X.}\ \bibnamefont
  {Liu}}, \bibinfo {author} {\bibfnamefont {J.~D.}\ \bibnamefont {Sau}},\ and\
  \bibinfo {author} {\bibfnamefont {S.}~\bibnamefont {Das~Sarma}},\ }\bibfield
  {title} {\bibinfo {title} {Role of dissipation in realistic Majorana
  nanowires},\ }\href {https://doi.org/10.1103/PhysRevB.95.054502} {\bibfield
  {journal} {\bibinfo  {journal} {Phys. Rev. B}\ }\textbf {\bibinfo {volume}
  {95}},\ \bibinfo {pages} {054502} (\bibinfo {year} {2017})}\BibitemShut
  {NoStop}%
\bibitem [{\citenamefont {Park}\ \emph {et~al.}(2011)\citenamefont {Park},
  \citenamefont {Kim}, \citenamefont {Yu}, \citenamefont {Han},\ and\
  \citenamefont {Kim}}]{park11}%
  \BibitemOpen
  \bibfield  {author} {\bibinfo {author} {\bibfnamefont {S.~R.}\ \bibnamefont
  {Park}}, \bibinfo {author} {\bibfnamefont {C.~H.}\ \bibnamefont {Kim}},
  \bibinfo {author} {\bibfnamefont {J.}~\bibnamefont {Yu}}, \bibinfo {author}
  {\bibfnamefont {J.~H.}\ \bibnamefont {Han}},\ and\ \bibinfo {author}
  {\bibfnamefont {C.}~\bibnamefont {Kim}},\ }\bibfield  {title} {\bibinfo
  {title} {Orbital-angular-momentum based origin of rashba-type surface band
  splitting},\ }\href@noop {} {\bibfield  {journal} {\bibinfo  {journal} {Phys.
  Rev. Lett.}\ }\textbf {\bibinfo {volume} {107}},\ \bibinfo {pages} {156803}
  (\bibinfo {year} {2011})}\BibitemShut {NoStop}%
\bibitem [{\citenamefont {Park}\ \emph {et~al.}(2012)\citenamefont {Park},
  \citenamefont {Kim}, \citenamefont {Rhim},\ and\ \citenamefont
  {Han}}]{park12}%
  \BibitemOpen
  \bibfield  {author} {\bibinfo {author} {\bibfnamefont {J.-H.}\ \bibnamefont
  {Park}}, \bibinfo {author} {\bibfnamefont {C.~H.}\ \bibnamefont {Kim}},
  \bibinfo {author} {\bibfnamefont {J.-W.}\ \bibnamefont {Rhim}},\ and\
  \bibinfo {author} {\bibfnamefont {J.~H.}\ \bibnamefont {Han}},\ }\bibfield
  {title} {\bibinfo {title} {Orbital Rashba effect and its detection by
  circular dichroism angle-resolved photoemission spectroscopy},\ }\href@noop
  {} {\bibfield  {journal} {\bibinfo  {journal} {Phys. Rev. B}\ }\textbf
  {\bibinfo {volume} {85}},\ \bibinfo {pages} {195401} (\bibinfo {year}
  {2012})}\BibitemShut {NoStop}%
\bibitem [{\citenamefont {Khalsa}\ \emph {et~al.}(2013)\citenamefont {Khalsa},
  \citenamefont {Lee},\ and\ \citenamefont {MacDonald}}]{Khalsa2013PRB}%
  \BibitemOpen
  \bibfield  {author} {\bibinfo {author} {\bibfnamefont {G.}~\bibnamefont
  {Khalsa}}, \bibinfo {author} {\bibfnamefont {B.}~\bibnamefont {Lee}},\ and\
  \bibinfo {author} {\bibfnamefont {A.~H.}\ \bibnamefont {MacDonald}},\
  }\bibfield  {title} {\bibinfo {title} {Theory of ${t}_{2g}$ electron-gas
  rashba interactions},\ }\href {https://doi.org/10.1103/PhysRevB.88.041302}
  {\bibfield  {journal} {\bibinfo  {journal} {Phys. Rev. B}\ }\textbf {\bibinfo
  {volume} {88}},\ \bibinfo {pages} {041302} (\bibinfo {year}
  {2013})}\BibitemShut {NoStop}%
\bibitem [{\citenamefont {Kim}\ \emph {et~al.}(2014)\citenamefont {Kim},
  \citenamefont {Kang}, \citenamefont {Go},\ and\ \citenamefont {Han}}]{kim14}%
  \BibitemOpen
  \bibfield  {author} {\bibinfo {author} {\bibfnamefont {P.}~\bibnamefont
  {Kim}}, \bibinfo {author} {\bibfnamefont {K.~T.}\ \bibnamefont {Kang}},
  \bibinfo {author} {\bibfnamefont {G.}~\bibnamefont {Go}},\ and\ \bibinfo
  {author} {\bibfnamefont {J.~H.}\ \bibnamefont {Han}},\ }\bibfield  {title}
  {\bibinfo {title} {Nature of orbital and spin rashba coupling in the surface
  bands of SrTiO$_3$ and KTaO$_3$},\ }\href@noop {} {\bibfield  {journal}
  {\bibinfo  {journal} {Phys. Rev. B}\ }\textbf {\bibinfo {volume} {90}},\
  \bibinfo {pages} {205423} (\bibinfo {year} {2014})}\BibitemShut {NoStop}%
\bibitem [{\citenamefont {Mercaldo}\ \emph {et~al.}(2020)\citenamefont
  {Mercaldo}, \citenamefont {Solinas}, \citenamefont {Giazotto},\ and\
  \citenamefont {Cuoco}}]{Mercaldo2020}%
  \BibitemOpen
  \bibfield  {author} {\bibinfo {author} {\bibfnamefont {M.~T.}\ \bibnamefont
  {Mercaldo}}, \bibinfo {author} {\bibfnamefont {P.}~\bibnamefont {Solinas}},
  \bibinfo {author} {\bibfnamefont {F.}~\bibnamefont {Giazotto}},\ and\
  \bibinfo {author} {\bibfnamefont {M.}~\bibnamefont {Cuoco}},\ }\bibfield
  {title} {\bibinfo {title} {Electrically tunable superconductivity through
  surface orbital polarization},\ }\href
  {https://doi.org/10.1103/PhysRevApplied.14.034041} {\bibfield  {journal}
  {\bibinfo  {journal} {Phys. Rev. Applied}\ }\textbf {\bibinfo {volume}
  {14}},\ \bibinfo {pages} {034041} (\bibinfo {year} {2020})}\BibitemShut
  {NoStop}%
\bibitem [{\citenamefont {Fukaya}\ \emph {et~al.}(2018)\citenamefont {Fukaya},
  \citenamefont {Tamura}, \citenamefont {Yada}, \citenamefont {Tanaka},
  \citenamefont {Gentile},\ and\ \citenamefont {Cuoco}}]{fukaya18}%
  \BibitemOpen
  \bibfield  {author} {\bibinfo {author} {\bibfnamefont {Y.}~\bibnamefont
  {Fukaya}}, \bibinfo {author} {\bibfnamefont {S.}~\bibnamefont {Tamura}},
  \bibinfo {author} {\bibfnamefont {K.}~\bibnamefont {Yada}}, \bibinfo {author}
  {\bibfnamefont {Y.}~\bibnamefont {Tanaka}}, \bibinfo {author} {\bibfnamefont
  {P.}~\bibnamefont {Gentile}},\ and\ \bibinfo {author} {\bibfnamefont
  {M.}~\bibnamefont {Cuoco}},\ }\bibfield  {title} {\bibinfo {title}
  {Interorbital topological superconductivity in spin-orbit coupled
  superconductors with inversion symmetry breaking},\ }\href
  {https://doi.org/10.1103/PhysRevB.97.174522} {\bibfield  {journal} {\bibinfo
  {journal} {Phys. Rev. B}\ }\textbf {\bibinfo {volume} {97}},\ \bibinfo
  {pages} {174522} (\bibinfo {year} {2018})}\BibitemShut {NoStop}%
\bibitem [{\citenamefont {Fukaya}\ \emph {et~al.}(2019)\citenamefont {Fukaya},
  \citenamefont {Tamura}, \citenamefont {Yada}, \citenamefont {Tanaka},
  \citenamefont {Gentile},\ and\ \citenamefont {Cuoco}}]{fukaya19}%
  \BibitemOpen
  \bibfield  {author} {\bibinfo {author} {\bibfnamefont {Y.}~\bibnamefont
  {Fukaya}}, \bibinfo {author} {\bibfnamefont {S.}~\bibnamefont {Tamura}},
  \bibinfo {author} {\bibfnamefont {K.}~\bibnamefont {Yada}}, \bibinfo {author}
  {\bibfnamefont {Y.}~\bibnamefont {Tanaka}}, \bibinfo {author} {\bibfnamefont
  {P.}~\bibnamefont {Gentile}},\ and\ \bibinfo {author} {\bibfnamefont
  {M.}~\bibnamefont {Cuoco}},\ }\bibfield  {title} {\bibinfo {title}
  {Spin-orbital hallmarks of unconventional superconductors without inversion
  symmetry},\ }\href {https://doi.org/10.1103/PhysRevB.100.104524} {\bibfield
  {journal} {\bibinfo  {journal} {Phys. Rev. B}\ }\textbf {\bibinfo {volume}
  {100}},\ \bibinfo {pages} {104524} (\bibinfo {year} {2019})}\BibitemShut
  {NoStop}%
\bibitem [{\citenamefont {Fukaya}\ \emph {et~al.}(2020)\citenamefont {Fukaya},
  \citenamefont {Yada}, \citenamefont {Tanaka}, \citenamefont {Gentile},\ and\
  \citenamefont {Cuoco}}]{fukaya20}%
  \BibitemOpen
  \bibfield  {author} {\bibinfo {author} {\bibfnamefont {Y.}~\bibnamefont
  {Fukaya}}, \bibinfo {author} {\bibfnamefont {K.}~\bibnamefont {Yada}},
  \bibinfo {author} {\bibfnamefont {Y.}~\bibnamefont {Tanaka}}, \bibinfo
  {author} {\bibfnamefont {P.}~\bibnamefont {Gentile}},\ and\ \bibinfo {author}
  {\bibfnamefont {M.}~\bibnamefont {Cuoco}},\ }\bibfield  {title} {\bibinfo
  {title} {Orbital tunable $0\ensuremath{-}\ensuremath{\pi}$ transitions in
  Josephson junctions with noncentrosymmetric topological superconductors},\
  }\href {https://doi.org/10.1103/PhysRevB.102.144512} {\bibfield  {journal}
  {\bibinfo  {journal} {Phys. Rev. B}\ }\textbf {\bibinfo {volume} {102}},\
  \bibinfo {pages} {144512} (\bibinfo {year} {2020})}\BibitemShut {NoStop}%
\bibitem [{\citenamefont {Sato}\ \emph {et~al.}(2011)\citenamefont {Sato},
  \citenamefont {Tanaka}, \citenamefont {Yada},\ and\ \citenamefont
  {Yokoyama}}]{Sato2011PRB}%
  \BibitemOpen
  \bibfield  {author} {\bibinfo {author} {\bibfnamefont {M.}~\bibnamefont
  {Sato}}, \bibinfo {author} {\bibfnamefont {Y.}~\bibnamefont {Tanaka}},
  \bibinfo {author} {\bibfnamefont {K.}~\bibnamefont {Yada}},\ and\ \bibinfo
  {author} {\bibfnamefont {T.}~\bibnamefont {Yokoyama}},\ }\bibfield  {title}
  {\bibinfo {title} {Topology of andreev bound states with flat dispersion},\
  }\href {https://doi.org/10.1103/PhysRevB.83.224511} {\bibfield  {journal}
  {\bibinfo  {journal} {Phys. Rev. B}\ }\textbf {\bibinfo {volume} {83}},\
  \bibinfo {pages} {224511} (\bibinfo {year} {2011})}\BibitemShut {NoStop}%
\bibitem [{\citenamefont {Yada}\ \emph {et~al.}(2011)\citenamefont {Yada},
  \citenamefont {Sato}, \citenamefont {Tanaka},\ and\ \citenamefont
  {Yokoyama}}]{Yada2011PRB}%
  \BibitemOpen
  \bibfield  {author} {\bibinfo {author} {\bibfnamefont {K.}~\bibnamefont
  {Yada}}, \bibinfo {author} {\bibfnamefont {M.}~\bibnamefont {Sato}}, \bibinfo
  {author} {\bibfnamefont {Y.}~\bibnamefont {Tanaka}},\ and\ \bibinfo {author}
  {\bibfnamefont {T.}~\bibnamefont {Yokoyama}},\ }\bibfield  {title} {\bibinfo
  {title} {Surface density of states and topological edge states in
  noncentrosymmetric superconductors},\ }\href
  {https://doi.org/10.1103/PhysRevB.83.064505} {\bibfield  {journal} {\bibinfo
  {journal} {Phys. Rev. B}\ }\textbf {\bibinfo {volume} {83}},\ \bibinfo
  {pages} {064505} (\bibinfo {year} {2011})}\BibitemShut {NoStop}%
\bibitem [{\citenamefont {Brydon}\ \emph {et~al.}(2011)\citenamefont {Brydon},
  \citenamefont {Schnyder},\ and\ \citenamefont {Timm}}]{Brydon2011PRB}%
  \BibitemOpen
  \bibfield  {author} {\bibinfo {author} {\bibfnamefont {P.~M.~R.}\
  \bibnamefont {Brydon}}, \bibinfo {author} {\bibfnamefont {A.~P.}\
  \bibnamefont {Schnyder}},\ and\ \bibinfo {author} {\bibfnamefont
  {C.}~\bibnamefont {Timm}},\ }\bibfield  {title} {\bibinfo {title}
  {Topologically protected flat zero-energy surface bands in noncentrosymmetric
  superconductors},\ }\href {https://doi.org/10.1103/PhysRevB.84.020501}
  {\bibfield  {journal} {\bibinfo  {journal} {Phys. Rev. B}\ }\textbf {\bibinfo
  {volume} {84}},\ \bibinfo {pages} {020501(R)} (\bibinfo {year}
  {2011})}\BibitemShut {NoStop}%
\bibitem [{\citenamefont {Mercaldo}\ \emph {et~al.}(2016)\citenamefont
  {Mercaldo}, \citenamefont {Cuoco},\ and\ \citenamefont
  {Kotetes}}]{Mercaldo16}%
  \BibitemOpen
  \bibfield  {author} {\bibinfo {author} {\bibfnamefont {M.~T.}\ \bibnamefont
  {Mercaldo}}, \bibinfo {author} {\bibfnamefont {M.}~\bibnamefont {Cuoco}},\
  and\ \bibinfo {author} {\bibfnamefont {P.}~\bibnamefont {Kotetes}},\
  }\bibfield  {title} {\bibinfo {title} {Magnetic-field-induced topological
  reorganization of a $p$-wave superconductor},\ }\href@noop {} {\bibfield
  {journal} {\bibinfo  {journal} {Phys. Rev. B}\ }\textbf {\bibinfo {volume}
  {94}},\ \bibinfo {pages} {140503(R)} (\bibinfo {year} {2016})}\BibitemShut
  {NoStop}%
\bibitem [{\citenamefont {Fukaya}\ \emph {et~al.}(2022)\citenamefont {Fukaya},
  \citenamefont {Tanaka}, \citenamefont {Gentile}, \citenamefont {Yada},\ and\
  \citenamefont {Cuoco}}]{Fukaya2022}%
  \BibitemOpen
  \bibfield  {author} {\bibinfo {author} {\bibfnamefont {Y.}~\bibnamefont
  {Fukaya}}, \bibinfo {author} {\bibfnamefont {Y.}~\bibnamefont {Tanaka}},
  \bibinfo {author} {\bibfnamefont {P.}~\bibnamefont {Gentile}}, \bibinfo
  {author} {\bibfnamefont {K.}~\bibnamefont {Yada}},\ and\ \bibinfo {author}
  {\bibfnamefont {M.}~\bibnamefont {Cuoco}},\ }\bibfield  {title} {\bibinfo
  {title} {Anomalous Josephson coupling and high-harmonics in
  non-centrosymmetric superconductors with s-wave spin-triplet pairing},\
  }\href {https://doi.org/10.1038/s41535-022-00509-8} {\bibfield  {journal}
  {\bibinfo  {journal} {npj Quantum Materials}\ }\textbf {\bibinfo {volume}
  {7}},\ \bibinfo {pages} {99} (\bibinfo {year} {2022})}\BibitemShut {NoStop}%
\bibitem [{Sup()}]{SupMat}%
  \BibitemOpen
  \bibfield  {title} {\bibinfo {title} {See Supplemental Material for the methodology to obtain the boundary Green's function. Then, we consider an effective two-orbital model with a pairing structure having orbital singlet symmetry and crystalline interactions that break the rotational symmetry and evaluate the density of states at the edge. We show that symmetric surface density of states generally occurs in single-orbital noncentrosymmetric superconductor with Rashba coupling} }\href@noop {\ }\BibitemShut {NoStop}%
\bibitem [{\citenamefont {Lee}\ and\ \citenamefont
  {Fisher}(1981)}]{Lee_FisherPRL1981}%
  \BibitemOpen
  \bibfield  {author} {\bibinfo {author} {\bibfnamefont {P.~A.}\ \bibnamefont
  {Lee}}\ and\ \bibinfo {author} {\bibfnamefont {D.~S.}\ \bibnamefont
  {Fisher}},\ }\bibfield  {title} {\bibinfo {title} {Anderson localization in
  two dimensions},\ }\href {https://doi.org/10.1103/PhysRevLett.47.882}
  {\bibfield  {journal} {\bibinfo  {journal} {Phys. Rev. Lett.}\ }\textbf
  {\bibinfo {volume} {47}},\ \bibinfo {pages} {882} (\bibinfo {year}
  {1981})}\BibitemShut {NoStop}%
\bibitem [{\citenamefont {Umerski}(1997)}]{LDOSUmerski}%
  \BibitemOpen
  \bibfield  {author} {\bibinfo {author} {\bibfnamefont {A.}~\bibnamefont
  {Umerski}},\ }\bibfield  {title} {\bibinfo {title} {Closed-form solutions to
  surface Green's functions},\ }\href
  {https://doi.org/10.1103/PhysRevB.55.5266} {\bibfield  {journal} {\bibinfo
  {journal} {Phys. Rev. B}\ }\textbf {\bibinfo {volume} {55}},\ \bibinfo
  {pages} {5266} (\bibinfo {year} {1997})}\BibitemShut {NoStop}%
\bibitem{KawaiPRB} K. Kawai, K. Yada, Y. Tanaka, Y. Asano, A. A. Golubov, and S. Kashiwaya, Josephson effect in a multiorbital model for
Sr$_2$RuO$_4$, Phys. Rev. B {\bf{95}}, 174518 (2017).
\bibitem{Tanaka2012} Y. Tanaka, M. Sato, and N. Nagaosa, Symmetry and
topology in superconductors –odd-frequency pairing and
edge states, J. Phys. Soc. Jpn. {\bf 81}, 011013 (2012).
\bibitem{Tanaka2007} Y. Tanaka and A. A. Golubov,
Theory of the Proximity Effect in Junctions with Unconventional Superconductors, Phys. Rev. Lett. {\bf 98}, 037003 (2007).
\bibitem{Tanaka2007a} Y. Tanaka, Y. Tanuma, and A. A. Golubov, Odd-frequency
pairing in normal-metal/superconductor junctions, Phys. Rev. B {\bf 76}, 054522 (2007).
\bibitem{Eschrig2007} M. Eschrig, T. Lofwander, T. Champel, J. C. Cuevas, J. Kopu, and G. Schon, Symmetries of pairing correlations in superconductor ferromagnet nanostructures, J. Low Temp. Phys. {\bf 147}, 457 (2007).
\bibitem{Linder2019} J. Linder and A. V. Balatsky, Odd-frequency superconductivity,
Rev. Mod. Phys. {\bf 91}, 045005 (2019).
\bibitem{Black2013} A. M. Black-Schaffer and A. V. Balatsky, Odd-frequency
superconducting pairing in multiband superconductors, Phys. Rev. B {\bf 88}, 104514 (2013).
\bibitem{Asano2015} Y. Asano and A. Sasaki, Odd-frequency cooper pairs in two-band superconductors and their magnetic response,
Phys. Rev. B {\bf 92}, 224508 (2015).
\bibitem{Ebisu2016} H. Ebisu, B. Lu, J. Klinovaja, and Y. Tanaka, Theory of time-reversal topological superconductivity in double Rashba wires: symmetries of Cooper pairs and Andreev bound states, Prog. Theor. Exp. Phys. 2016, (2016).
\bibitem [{\citenamefont {Miao}\ \emph {et~al.}(2020)\citenamefont {Miao},
  \citenamefont {Liu}, \citenamefont {Xu}, \citenamefont {Kotta}, \citenamefont
  {Kang}, \citenamefont {Ran}, \citenamefont {Paglione}, \citenamefont
  {Kotliar}, \citenamefont {Butch}, \citenamefont {Denlinger},\ and\
  \citenamefont {Wray}}]{Lin2020}%
  \BibitemOpen
  \bibfield  {author} {\bibinfo {author} {\bibfnamefont {L.}~\bibnamefont
  {Miao}}, \bibinfo {author} {\bibfnamefont {S.}~\bibnamefont {Liu}}, \bibinfo
  {author} {\bibfnamefont {Y.}~\bibnamefont {Xu}}, \bibinfo {author}
  {\bibfnamefont {E.~C.}\ \bibnamefont {Kotta}}, \bibinfo {author}
  {\bibfnamefont {C.-J.}\ \bibnamefont {Kang}}, \bibinfo {author}
  {\bibfnamefont {S.}~\bibnamefont {Ran}}, \bibinfo {author} {\bibfnamefont
  {J.}~\bibnamefont {Paglione}}, \bibinfo {author} {\bibfnamefont
  {G.}~\bibnamefont {Kotliar}}, \bibinfo {author} {\bibfnamefont {N.~P.}\
  \bibnamefont {Butch}}, \bibinfo {author} {\bibfnamefont {J.~D.}\ \bibnamefont
  {Denlinger}},\ and\ \bibinfo {author} {\bibfnamefont {L.~A.}\ \bibnamefont
  {Wray}},\ }\bibfield  {title} {\bibinfo {title} {Low energy band structure
  and symmetries of ${\mathrm{UTe}}_{2}$ from angle-resolved photoemission
  spectroscopy},\ }\href {https://doi.org/10.1103/PhysRevLett.124.076401}
  {\bibfield  {journal} {\bibinfo  {journal} {Phys. Rev. Lett.}\ }\textbf
  {\bibinfo {volume} {124}},\ \bibinfo {pages} {076401} (\bibinfo {year}
  {2020})}\BibitemShut {NoStop}%
\bibitem [{\citenamefont {Ohtomo}\ and\ \citenamefont
  {Hwang}(2004)}]{Ohtomo-2004}%
  \BibitemOpen
  \bibfield  {author} {\bibinfo {author} {\bibfnamefont {A.}~\bibnamefont
  {Ohtomo}}\ and\ \bibinfo {author} {\bibfnamefont {H.~Y.}\ \bibnamefont
  {Hwang}},\ }\bibfield  {title} {\bibinfo {title} {A high-mobility electron
  gas at the LaAlO$_3$/SrTiO$_3$ heterointerface},\ }\href
  {https://doi.org/10.1038/nature02308} {\bibfield  {journal} {\bibinfo
  {journal} {Nature}\ }\textbf {\bibinfo {volume} {427}},\ \bibinfo {pages}
  {423} (\bibinfo {year} {2004})}\BibitemShut {NoStop}%
\bibitem [{\citenamefont {Reyren}\ \emph {et~al.}(2007)\citenamefont {Reyren},
  \citenamefont {Thiel}, \citenamefont {Caviglia}, \citenamefont {Kourkoutis},
  \citenamefont {Hammerl}, \citenamefont {Richter}, \citenamefont {Schneider},
  \citenamefont {Kopp}, \citenamefont {Ruetschi}, \citenamefont {Jaccard},
  \citenamefont {Gabay}, \citenamefont {Muller}, \citenamefont {Triscone},
  \citenamefont {J.Mannhart}, \citenamefont {Kourkoutis},\ and\ \citenamefont
  {Hammerl}}]{Reyren-2007}%
  \BibitemOpen
  \bibfield  {author} {\bibinfo {author} {\bibfnamefont {N.}~\bibnamefont
  {Reyren}}, \bibinfo {author} {\bibfnamefont {S.}~\bibnamefont {Thiel}},
  \bibinfo {author} {\bibfnamefont {A.~D.}\ \bibnamefont {Caviglia}}, \bibinfo
  {author} {\bibfnamefont {L.~F.}\ \bibnamefont {Kourkoutis}}, \bibinfo
  {author} {\bibfnamefont {G.}~\bibnamefont {Hammerl}}, \bibinfo {author}
  {\bibfnamefont {C.}~\bibnamefont {Richter}}, \bibinfo {author} {\bibfnamefont
  {C.}~\bibnamefont {Schneider}}, \bibinfo {author} {\bibfnamefont
  {T.}~\bibnamefont {Kopp}}, \bibinfo {author} {\bibfnamefont {A.-S.}\
  \bibnamefont {Ruetschi}}, \bibinfo {author} {\bibfnamefont {D.}~\bibnamefont
  {Jaccard}}, \bibinfo {author} {\bibfnamefont {M.}~\bibnamefont {Gabay}},
  \bibinfo {author} {\bibfnamefont {D.~A.}\ \bibnamefont {Muller}}, \bibinfo
  {author} {\bibfnamefont {J.-M.~M.}\ \bibnamefont {Triscone}}, \bibinfo
  {author} {\bibnamefont {J.Mannhart}}, \bibinfo {author} {\bibfnamefont
  {L.~F.}\ \bibnamefont {Kourkoutis}},\ and\ \bibinfo {author} {\bibfnamefont
  {G.}~\bibnamefont {Hammerl}},\ }\bibfield  {title} {\bibinfo {title}
  {Superconducting interfaces between insulating oxides},\ }\href
  {https://doi.org/10.1126/science.1146006} {\bibfield  {journal} {\bibinfo
  {journal} {Science}\ }\textbf {\bibinfo {volume} {317}},\ \bibinfo {pages}
  {1196} (\bibinfo {year} {2007})}\BibitemShut {NoStop}%
\bibitem [{\citenamefont {Caviglia}\ and\ \citenamefont
  {et~al.}(2008)}]{Caviglia-2008}%
  \BibitemOpen
  \bibfield  {author} {\bibinfo {author} {\bibfnamefont {A.~D.}\ \bibnamefont
  {Caviglia,}} \bibinfo {author} {\bibnamefont {S. Gariglio, N. Reyren, D. Jaccard, T. Schneider, M. Gabay, S. Thiel, G. Hammerl, J. Mannhart, and J.-M. Triscone}},\ }\bibfield
  {title} {\bibinfo {title} {Electric field control of the LaAlO$_3$/SrTiO$_3$
  interface ground state},\ }\href {https://doi.org/10.1038/nature07576}
  {\bibfield  {journal} {\bibinfo  {journal} {Nature}\ }\textbf {\bibinfo
  {volume} {456}},\ \bibinfo {pages} {624} (\bibinfo {year}
  {2008})}\BibitemShut {NoStop}%
\bibitem [{\citenamefont {Thierschmann}\ and\ \citenamefont
  {et~al.}(2018)}]{Thierschmann-2018}%
  \BibitemOpen
  \bibfield  {author} {\bibinfo {author} {\bibfnamefont {H.}~\bibnamefont
  {Thierschmann,}}\bibinfo {author} \, {\bibnamefont {E. Mulazimoglu, N. Manca, S. Goswami, T. M. Klapwijk, and A. Caviglia }},\ }\bibfield
  {title} {\bibinfo {title} {Transport regimes of a split gate superconducting
  quantum point contact in the two-dimensional LaAlO$_3$/SrTiO$_3$
  superfluid},\ }\href {https://doi.org/10.1038/s41467-018-04657-z} {\bibfield
  {journal} {\bibinfo  {journal} {Nat. Commun.}\ }\textbf {\bibinfo {volume}
  {9}},\ \bibinfo {pages} {2276} (\bibinfo {year} {2018})}\BibitemShut
  {NoStop}%
\bibitem [{\citenamefont {Hurand}\ and\ \citenamefont
  {et~al.}(2015)}]{Hurand-2015}%
  \BibitemOpen
  \bibfield  {author} {\bibinfo {author} {\bibfnamefont {S.}~\bibnamefont
  {Hurand,}}\ \ \bibinfo {author} {\bibnamefont {A. Jouan, C. Feuillet-Palma, G. Singh, J. Biscaras, E. Lesne, N. Reyren, A. Barthélémy, M. Bibes, J. E. Villegas, C. Ulysse, X. Lafosse, M. Pannetier-Lecoeur, S. Caprara, M. Grilli, J. Lesueur, and N. Bergeal}},\ }\bibfield
  {title} {\bibinfo {title} {Field-effect control of superconductivity and
  rashba spin–orbit coupling in top-gated LaAlO$_3$/SrTiO$_3$ devices},\
  }\href {https://doi.org/10.1038/srep12751} {\bibfield  {journal} {\bibinfo
  {journal} {Sci. Rep.}\ }\textbf {\bibinfo {volume} {5}},\ \bibinfo {pages}
  {12751} (\bibinfo {year} {2015})}\BibitemShut {NoStop}%
\bibitem [{\citenamefont {Caviglia}\ \emph {et~al.}(2010)\citenamefont
  {Caviglia}, \citenamefont {Gabay}, \citenamefont {Gariglio}, \citenamefont
  {Reyren}, \citenamefont {Cancellieri},\ and\ \citenamefont
  {Triscone}}]{Caviglia2010PRL}%
  \BibitemOpen
  \bibfield  {author} {\bibinfo {author} {\bibfnamefont {A.~D.}\ \bibnamefont
  {Caviglia}}, \bibinfo {author} {\bibfnamefont {M.}~\bibnamefont {Gabay}},
  \bibinfo {author} {\bibfnamefont {S.}~\bibnamefont {Gariglio}}, \bibinfo
  {author} {\bibfnamefont {N.}~\bibnamefont {Reyren}}, \bibinfo {author}
  {\bibfnamefont {C.}~\bibnamefont {Cancellieri}},\ and\ \bibinfo {author}
  {\bibfnamefont {J.-M.}\ \bibnamefont {Triscone}},\ }\bibfield  {title}
  {\bibinfo {title} {Tunable rashba spin-orbit interaction at oxide
  interfaces},\ }\href {https://doi.org/10.1103/PhysRevLett.104.126803}
  {\bibfield  {journal} {\bibinfo  {journal} {Phys. Rev. Lett.}\ }\textbf
  {\bibinfo {volume} {104}},\ \bibinfo {pages} {126803} (\bibinfo {year}
  {2010})}\BibitemShut {NoStop}%
\bibitem [{\citenamefont {Ben~Shalom}\ \emph {et~al.}(2010)\citenamefont
  {Ben~Shalom}, \citenamefont {Sachs}, \citenamefont {Rakhmilevitch},
  \citenamefont {Palevski},\ and\ \citenamefont {Dagan}}]{Ben_Shalom2010PRL}%
  \BibitemOpen
  \bibfield  {author} {\bibinfo {author} {\bibfnamefont {M.}~\bibnamefont
  {Ben~Shalom}}, \bibinfo {author} {\bibfnamefont {M.}~\bibnamefont {Sachs}},
  \bibinfo {author} {\bibfnamefont {D.}~\bibnamefont {Rakhmilevitch}}, \bibinfo
  {author} {\bibfnamefont {A.}~\bibnamefont {Palevski}},\ and\ \bibinfo
  {author} {\bibfnamefont {Y.}~\bibnamefont {Dagan}},\ }\bibfield  {title}
  {\bibinfo {title} {Tuning spin-orbit coupling and superconductivity at the
  SrTiO$_{3}$/LaAlO$_{3}$ interface: A magnetotransport
  study},\ }\href {https://doi.org/10.1103/PhysRevLett.104.126802} {\bibfield
  {journal} {\bibinfo  {journal} {Phys. Rev. Lett.}\ }\textbf {\bibinfo
  {volume} {104}},\ \bibinfo {pages} {126802} (\bibinfo {year}
  {2010})}\BibitemShut {NoStop}%
\bibitem [{\citenamefont {Joshua}\ \emph {et~al.}(2012)\citenamefont {Joshua},
  \citenamefont {Pecker}, \citenamefont {Ruhman}, \citenamefont {Altman},\ and\
  \citenamefont {Ilani}}]{Joshua-2012}%
  \BibitemOpen
  \bibfield  {author} {\bibinfo {author} {\bibfnamefont {A.}~\bibnamefont
  {Joshua}}, \bibinfo {author} {\bibfnamefont {S.}~\bibnamefont {Pecker}},
  \bibinfo {author} {\bibfnamefont {J.}~\bibnamefont {Ruhman}}, \bibinfo
  {author} {\bibfnamefont {E.}~\bibnamefont {Altman}},\ and\ \bibinfo {author}
  {\bibfnamefont {S.}~\bibnamefont {Ilani}},\ }\bibfield  {title} {\bibinfo
  {title} {A universal critical density underlying the physics of electrons at
  the laalo$_3$/srtio$_3$ interface},\ }\href
  {https://doi.org/10.1038/ncomms2116} {\bibfield  {journal} {\bibinfo
  {journal} {Nat. Commun.}\ }\textbf {\bibinfo {volume} {3}},\ \bibinfo {pages}
  {1129} (\bibinfo {year} {2012})}\BibitemShut {NoStop}%
\bibitem [{\citenamefont {Herranz}\ \emph {et~al.}(2015)\citenamefont
  {Herranz}, \citenamefont {Singh}, \citenamefont {Bergeal}, \citenamefont
  {Jouan}, \citenamefont {Lesueur}, \citenamefont {Gázquez}, \citenamefont
  {Varela}, \citenamefont {Scigaj}, \citenamefont {Dix}, \citenamefont
  {S\'{a}nchez},\ and\ \citenamefont {Fontcuberta}}]{Herranz-2015}%
  \BibitemOpen
  \bibfield  {author} {\bibinfo {author} {\bibfnamefont {G.}~\bibnamefont
  {Herranz}}, \bibinfo {author} {\bibfnamefont {G.}~\bibnamefont {Singh}},
  \bibinfo {author} {\bibfnamefont {N.}~\bibnamefont {Bergeal}}, \bibinfo
  {author} {\bibfnamefont {A.}~\bibnamefont {Jouan}}, \bibinfo {author}
  {\bibfnamefont {J.}~\bibnamefont {Lesueur}}, \bibinfo {author} {\bibfnamefont
  {J.}~\bibnamefont {Gázquez}}, \bibinfo {author} {\bibfnamefont
  {M.}~\bibnamefont {Varela}}, \bibinfo {author} {\bibfnamefont
  {M.}~\bibnamefont {Scigaj}}, \bibinfo {author} {\bibfnamefont
  {N.}~\bibnamefont {Dix}}, \bibinfo {author} {\bibfnamefont {F.}~\bibnamefont
  {S\'{a}nchez}},\ and\ \bibinfo {author} {\bibfnamefont {J.}~\bibnamefont
  {Fontcuberta}},\ }\bibfield  {title} {\bibinfo {title} {Engineering
  two-dimensional superconductivity and rashba spin–orbit coupling in
  LaAlO$_3$/SrTiO$_3$ quantum wells by selective orbital occupancy},\ }\href
  {https://doi.org/10.1038/ncomms7028} {\bibfield  {journal} {\bibinfo
  {journal} {Nat. Commun.}\ }\textbf {\bibinfo {volume} {6}},\ \bibinfo {pages}
  {6028} (\bibinfo {year} {2015})}\BibitemShut {NoStop}%
\bibitem{Lesne2023}  E. Lesne, Y. G. Sa{\v g}lam, R. Battilomo, M. T. Mercaldo, T. C. van Thiel, U. Filippozzi, C. Noce, M. Cuoco, G. A. Steele, C. Ortix, A. D. Caviglia, Designing spin and orbital sources of Berry curvature at oxide interfaces, Nature Materials {\bf 22}, 576 (2023).
\bibitem{Mercaldo2023} M. T. Mercaldo, C. Noce, A. D. Caviglia, M. Cuoco, C. Ortix, Orbital design of Berry curvature: pinch points and giant dipoles induced by crystal fields, npj Quantum Materials {\bf 8}, 12 (2023).
\bibitem [{\citenamefont {Singh}\ \emph {et~al.}(2019)\citenamefont {Singh},
  \citenamefont {Jouan}, \citenamefont {Herranz}, \citenamefont {Scigaj},
  \citenamefont {S\'{a}nchez}, \citenamefont {Benfatto}, \citenamefont
  {Caprara}, \citenamefont {Grilli}, \citenamefont {Saiz}, \citenamefont
  {Couedo}, \citenamefont {Feuillet-Palma}, \citenamefont {Lesueur},\ and\
  \citenamefont {Bergeal}}]{Singh-2019}%
  \BibitemOpen
  \bibfield  {author} {\bibinfo {author} {\bibfnamefont {G.}~\bibnamefont
  {Singh}}, \bibinfo {author} {\bibfnamefont {A.}~\bibnamefont {Jouan}},
  \bibinfo {author} {\bibfnamefont {G.}~\bibnamefont {Herranz}}, \bibinfo
  {author} {\bibfnamefont {M.}~\bibnamefont {Scigaj}}, \bibinfo {author}
  {\bibfnamefont {F.}~\bibnamefont {S\'{a}nchez}}, \bibinfo {author}
  {\bibfnamefont {L.}~\bibnamefont {Benfatto}}, \bibinfo {author}
  {\bibfnamefont {S.}~\bibnamefont {Caprara}}, \bibinfo {author} {\bibfnamefont
  {M.}~\bibnamefont {Grilli}}, \bibinfo {author} {\bibfnamefont
  {G.}~\bibnamefont {Saiz}}, \bibinfo {author} {\bibfnamefont {F.}~\bibnamefont
  {Couedo}}, \bibinfo {author} {\bibfnamefont {C.}~\bibnamefont
  {Feuillet-Palma}}, \bibinfo {author} {\bibfnamefont {J.}~\bibnamefont
  {Lesueur}},\ and\ \bibinfo {author} {\bibfnamefont {N.}~\bibnamefont
  {Bergeal}},\ }\bibfield  {title} {\bibinfo {title} {Gap suppression at a
  Lifshitz transition in a multi-condensate superconductor},\ }\href
  {https://doi.org/10.1038/s41563-019-0354-z} {\bibfield  {journal} {\bibinfo
  {journal} {Nat. Mater.}\ }\textbf {\bibinfo {volume} {18}},\ \bibinfo {pages}
  {948} (\bibinfo {year} {2019})}\BibitemShut {NoStop}%
\bibitem [{\citenamefont {Trevisan}\ \emph {et~al.}(2018)\citenamefont
  {Trevisan}, \citenamefont {Sch\"utt},\ and\ \citenamefont
  {Fernandes}}]{Trevisan-2018}%
  \BibitemOpen
  \bibfield  {author} {\bibinfo {author} {\bibfnamefont {T.~V.}\ \bibnamefont
  {Trevisan}}, \bibinfo {author} {\bibfnamefont {M.}~\bibnamefont {Sch\"utt}},\
  and\ \bibinfo {author} {\bibfnamefont {R.~M.}\ \bibnamefont {Fernandes}},\
  }\bibfield  {title} {\bibinfo {title} {Unconventional multiband
  superconductivity in bulk ${\mathrm{srtio}}_{3}$ and
  ${\mathrm{laalo}}_{3}/{\mathrm{srtio}}_{3}$ interfaces},\ }\href
  {https://doi.org/10.1103/PhysRevLett.121.127002} {\bibfield  {journal}
  {\bibinfo  {journal} {Phys. Rev. Lett.}\ }\textbf {\bibinfo {volume} {121}},\
  \bibinfo {pages} {127002} (\bibinfo {year} {2018})}\BibitemShut {NoStop}%
\bibitem [{\citenamefont {Bal}\ \emph {et~al.}(2015)\citenamefont {Bal},
  \citenamefont {Mehta}, \citenamefont {Ryu}, \citenamefont {Lee},
  \citenamefont {Folkman}, \citenamefont {Eom},\ and\ \citenamefont
  {Chandrasekhar}}]{Bal-2015}%
  \BibitemOpen
  \bibfield  {author} {\bibinfo {author} {\bibfnamefont {V.~V.}\ \bibnamefont
  {Bal}}, \bibinfo {author} {\bibfnamefont {M.~M.}\ \bibnamefont {Mehta}},
  \bibinfo {author} {\bibfnamefont {S.}~\bibnamefont {Ryu}}, \bibinfo {author}
  {\bibfnamefont {H.}~\bibnamefont {Lee}}, \bibinfo {author} {\bibfnamefont
  {C.~M.}\ \bibnamefont {Folkman}}, \bibinfo {author} {\bibfnamefont {C.~B.}\
  \bibnamefont {Eom}},\ and\ \bibinfo {author} {\bibfnamefont {V.}~\bibnamefont
  {Chandrasekhar}},\ }\bibfield  {title} {\bibinfo {title} {Gate-tunable
  superconducting weak link behavior in top-gated
  LaAlO$_3$-SrTiO$_3$},\ }\href
  {https://doi.org/10.1063/1.4921924} {\bibfield  {journal} {\bibinfo
  {journal} {Appl. Phys. Lett.}\ }\textbf {\bibinfo {volume} {106}},\ \bibinfo
  {pages} {212601} (\bibinfo {year} {2015})}\BibitemShut {NoStop}%
\bibitem [{\citenamefont {Stornaiuolo}\ \emph {et~al.}(2017)\citenamefont
  {Stornaiuolo}, \citenamefont {Massarotti}, \citenamefont {Capua},
  \citenamefont {Lucignano}, \citenamefont {Pepe}, \citenamefont {Salluzzo},\
  and\ \citenamefont {Tafuri}}]{Stornaiuolo2017}%
  \BibitemOpen
  \bibfield  {author} {\bibinfo {author} {\bibfnamefont {D.}~\bibnamefont
  {Stornaiuolo}}, \bibinfo {author} {\bibfnamefont {D.}~\bibnamefont
  {Massarotti}}, \bibinfo {author} {\bibfnamefont {R.~D.}\ \bibnamefont
  {Capua}}, \bibinfo {author} {\bibfnamefont {P.}~\bibnamefont {Lucignano}},
  \bibinfo {author} {\bibfnamefont {G.~P.}\ \bibnamefont {Pepe}}, \bibinfo
  {author} {\bibfnamefont {M.}~\bibnamefont {Salluzzo}},\ and\ \bibinfo
  {author} {\bibfnamefont {F.}~\bibnamefont {Tafuri}},\ }\bibfield  {title}
  {\bibinfo {title} {Signatures of unconventional superconductivity in the
  LaAlO$_3$/SrTiO$_3$ two-dimensional system},\ }\href
  {https://doi.org/https://doi.org/10.1103/PhysRevB.95.140502} {\bibfield
  {journal} {\bibinfo  {journal} {Phys. Rev. B}\ }\textbf {\bibinfo {volume}
  {95}},\ \bibinfo {pages} {140502(R)} (\bibinfo {year} {2017})}\BibitemShut
  {NoStop}%
\bibitem [{\citenamefont {Kalaboukhov}\ \emph {et~al.}(2017)\citenamefont
  {Kalaboukhov}, \citenamefont {Aurino}, \citenamefont {Galletti},
  \citenamefont {Bauch}, \citenamefont {Lombardi}, \citenamefont {Winkler},
  \citenamefont {Claeson},\ and\ \citenamefont {Golubev}}]{Kalaboukhov-2017}%
  \BibitemOpen
  \bibfield  {author} {\bibinfo {author} {\bibfnamefont {A.}~\bibnamefont
  {Kalaboukhov}}, \bibinfo {author} {\bibfnamefont {P.~P.}\ \bibnamefont
  {Aurino}}, \bibinfo {author} {\bibfnamefont {L.}~\bibnamefont {Galletti}},
  \bibinfo {author} {\bibfnamefont {T.}~\bibnamefont {Bauch}}, \bibinfo
  {author} {\bibfnamefont {F.}~\bibnamefont {Lombardi}}, \bibinfo {author}
  {\bibfnamefont {D.}~\bibnamefont {Winkler}}, \bibinfo {author} {\bibfnamefont
  {T.}~\bibnamefont {Claeson}},\ and\ \bibinfo {author} {\bibfnamefont
  {D.}~\bibnamefont {Golubev}},\ }\bibfield  {title} {\bibinfo {title}
  {Homogeneous superconductivity at the
  ${\mathrm{laalo}}_{3}/{\mathrm{srtio}}_{3}$ interface probed by nanoscale
  transport},\ }\href {https://doi.org/10.1103/PhysRevB.96.184525} {\bibfield
  {journal} {\bibinfo  {journal} {Phys. Rev. B}\ }\textbf {\bibinfo {volume}
  {96}},\ \bibinfo {pages} {184525} (\bibinfo {year} {2017})}\BibitemShut
  {NoStop}%
\bibitem [{\citenamefont {Kuerten}\ \emph {et~al.}(2017)\citenamefont
  {Kuerten}, \citenamefont {Richter}, \citenamefont {Mohanta}, \citenamefont
  {Kopp}, \citenamefont {Kampf}, \citenamefont {Mannhart},\ and\ \citenamefont
  {Boschker}}]{Kuerten-2017}%
  \BibitemOpen
  \bibfield  {author} {\bibinfo {author} {\bibfnamefont {L.}~\bibnamefont
  {Kuerten}}, \bibinfo {author} {\bibfnamefont {C.}~\bibnamefont {Richter}},
  \bibinfo {author} {\bibfnamefont {N.}~\bibnamefont {Mohanta}}, \bibinfo
  {author} {\bibfnamefont {T.}~\bibnamefont {Kopp}}, \bibinfo {author}
  {\bibfnamefont {A.}~\bibnamefont {Kampf}}, \bibinfo {author} {\bibfnamefont
  {J.}~\bibnamefont {Mannhart}},\ and\ \bibinfo {author} {\bibfnamefont
  {H.}~\bibnamefont {Boschker}},\ }\bibfield  {title} {\bibinfo {title} {In-gap
  states in superconducting
  ${\mathrm{LaAlO}}_{3}\ensuremath{-}{\mathrm{SrTiO}}_{3}$ interfaces observed
  by tunneling spectroscopy},\ }\href
  {https://doi.org/10.1103/PhysRevB.96.014513} {\bibfield  {journal} {\bibinfo
  {journal} {Phys. Rev. B}\ }\textbf {\bibinfo {volume} {96}},\ \bibinfo
  {pages} {014513} (\bibinfo {year} {2017})}\BibitemShut {NoStop}%
\bibitem [{\citenamefont {Singh}\ \emph {et~al.}(2022)\citenamefont {Singh},
  \citenamefont {Guarcello}, \citenamefont {Lesne}, \citenamefont {Winkler},
  \citenamefont {Claeson}, \citenamefont {Bauch}, \citenamefont {Lombardi},
  \citenamefont {Caviglia}, \citenamefont {Citro}, \citenamefont {Cuoco},\ and\
  \citenamefont {Kalaboukhov}}]{Sin21}%
  \BibitemOpen
  \bibfield  {author} {\bibinfo {author} {\bibfnamefont {G.}~\bibnamefont
  {Singh}}, \bibinfo {author} {\bibfnamefont {C.}~\bibnamefont {Guarcello}},
  \bibinfo {author} {\bibfnamefont {E.}~\bibnamefont {Lesne}}, \bibinfo
  {author} {\bibfnamefont {D.}~\bibnamefont {Winkler}}, \bibinfo {author}
  {\bibfnamefont {T.}~\bibnamefont {Claeson}}, \bibinfo {author} {\bibfnamefont
  {T.}~\bibnamefont {Bauch}}, \bibinfo {author} {\bibfnamefont
  {F.}~\bibnamefont {Lombardi}}, \bibinfo {author} {\bibfnamefont {A.~D.}\
  \bibnamefont {Caviglia}}, \bibinfo {author} {\bibfnamefont {R.}~\bibnamefont
  {Citro}}, \bibinfo {author} {\bibfnamefont {M.}~\bibnamefont {Cuoco}},\ and\
  \bibinfo {author} {\bibfnamefont {A.}~\bibnamefont {Kalaboukhov}},\
  }\bibfield  {title} {\bibinfo {title} {Gate-tunable pairing channels in
  superconducting non-centrosymmetric oxides nanowires},\ }\href
  {https://doi.org/https://doi.org/10.1038/s41535-021-00406-6} {\bibfield
  {journal} {\bibinfo  {journal} {npj Quantum Materials}\ }\textbf {\bibinfo
  {volume} {7}},\ \bibinfo {pages} {2} (\bibinfo {year} {2022})}\BibitemShut
  {NoStop}%
\bibitem [{\citenamefont {Mizushima}\ \emph {et~al.}(2014)\citenamefont
  {Mizushima}, \citenamefont {Yamakage}, \citenamefont {Sato},\ and\
  \citenamefont {Tanaka}}]{Mizushima2014}%
  \BibitemOpen
  \bibfield  {author} {\bibinfo {author} {\bibfnamefont {T.}~\bibnamefont
  {Mizushima}}, \bibinfo {author} {\bibfnamefont {A.}~\bibnamefont {Yamakage}},
  \bibinfo {author} {\bibfnamefont {M.}~\bibnamefont {Sato}},\ and\ \bibinfo
  {author} {\bibfnamefont {Y.}~\bibnamefont {Tanaka}},\ }\href
  {https://doi.org/10.1103/PhysRevB.90.184516} {\bibfield  {journal} {\bibinfo
  {journal} {Dirac-fermion-induced parity mixing in superconducting topological insulators, Phys. Rev. B}\ }\textbf {\bibinfo {volume} {90}},\ \bibinfo
  {pages} {184516} (\bibinfo {year} {2014})}\BibitemShut {NoStop}%
\bibitem [{\citenamefont {Fu}\ and\ \citenamefont {Berg}(2010)}]{Fu_Berg_2010}%
  \BibitemOpen
  \bibfield  {author} {\bibinfo {author} {\bibfnamefont {L.}~\bibnamefont
  {Fu}}\ and\ \bibinfo {author} {\bibfnamefont {E.}~\bibnamefont {Berg}},\
  }\bibfield  {title} {\bibinfo {title} {Odd-parity topological
  superconductors: Theory and application to
  ${\mathrm{Cu}}_{x}{\mathrm{Bi}}_{2}{\mathrm{Se}}_{3}$},\ }\href
  {https://doi.org/10.1103/PhysRevLett.105.097001} {\bibfield  {journal}
  {\bibinfo  {journal} {Phys. Rev. Lett.}\ }\textbf {\bibinfo {volume} {105}},\
  \bibinfo {pages} {097001} (\bibinfo {year} {2010})}\BibitemShut {NoStop}%
\bibitem [{\citenamefont {Nakosai}\ \emph {et~al.}(2012)\citenamefont
  {Nakosai}, \citenamefont {Tanaka},\ and\ \citenamefont
  {Nagaosa}}]{Nakosai2012}%
  \BibitemOpen
  \bibfield  {author} {\bibinfo {author} {\bibfnamefont {S.}~\bibnamefont
  {Nakosai}}, \bibinfo {author} {\bibfnamefont {Y.}~\bibnamefont {Tanaka}},\
  and\ \bibinfo {author} {\bibfnamefont {N.}~\bibnamefont {Nagaosa}},\
  }\bibfield  {title} {\bibinfo {title} {Topological superconductivity in
  bilayer Rashba system},\ }\href
  {https://doi.org/10.1103/PhysRevLett.108.147003} {\bibfield  {journal}
  {\bibinfo  {journal} {Phys. Rev. Lett.}\ }\textbf {\bibinfo {volume} {108}},\
  \bibinfo {pages} {147003} (\bibinfo {year} {2012})}\BibitemShut {NoStop}%
\bibitem [{\citenamefont {Xu}\ \emph {et~al.}(2014)\citenamefont {Xu},
  \citenamefont {Yao}, \citenamefont {Xiao},\ and\ \citenamefont
  {Heinz}}]{Xu2014}%
  \BibitemOpen
  \bibfield  {author} {\bibinfo {author} {\bibfnamefont {X.}~\bibnamefont
  {Xu}}, \bibinfo {author} {\bibfnamefont {W.}~\bibnamefont {Yao}}, \bibinfo
  {author} {\bibfnamefont {D.}~\bibnamefont {Xiao}},\ and\ \bibinfo {author}
  {\bibfnamefont {T.~F.}\ \bibnamefont {Heinz}},\ }\bibfield  {title} {\bibinfo
  {title} {Spin and pseudospins in layered transition metal dichalcogenides},\
  }\href {https://doi.org/10.1038/nphys2942} {\bibfield  {journal} {\bibinfo
  {journal} {Nature Physics}\ }\textbf {\bibinfo {volume} {10}},\ \bibinfo
  {pages} {343} (\bibinfo {year} {2014})}\BibitemShut {NoStop}%
\bibitem{Mutch2019} J. Mutch, W.-C. Chen, P. Went, T. Qian, I. Z. Wilson, A. Andreev, C.-C. Chen, and J.-H. Chu, Evidence for a strain-tuned topological phase transition in 
ZrTe$_5$, Sci. Adv. {\bf 5}, eaav9771 (2019).
\bibitem{Simoni2018} G. De Simoni, F. Paolucci, P. Solinas, E. Strambini, F. Giazotto, Metallic supercurrent field-effect transistor, Nature Nanotechnology 13, 802 (2018).
\bibitem{Simoni2021} G. De Simoni, S. Battisti, N. Ligato, M. T. Mercaldo, M. Cuoco, and F. Giazotto, Gate Control of the Current–Flux Relation of a Josephson Quantum Interferometer Based on Proximitized Metallic Nanojuntions, ACS Appl. Electron. Mater. {\bf 3}, 3927 (2021).
\bibitem{Mercaldo2021} M. T. Mercaldo, F. Giazotto, and M. Cuoco, Spectroscopic signatures of gate-controlled superconducting phases, Phys. Rev. Research {\bf 3}, 043042 (2021). 
\bibitem{MKC2019} M. T. Mercaldo, P. Kotetes, and M. Cuoco, Magnetoelectrically tunable Andreev bound state spectra and spin polarization in p-wave Josephson junctions, Phys. Rev. B {\bf 100}, 104519 (2019).
\bibitem{MKC2018} M. T. Mercaldo, P. Kotetes, and M. Cuoco, Magnetic manipulation of topological states in p-wave superconductors, Physica B {\bf 536}, 730 (2018).
\bibitem{Chirolli} L. Chirolli, M. T. Mercaldo, C. Guarcello, F. Giazotto, and M. Cuoco, Colossal Orbital Edelstein Effect in Noncentrosymmetric Superconductors,
Phys. Rev. Lett. {\bf 128}, 217703 (2022).
\end{thebibliography}

%

\end{document}


\title{Particle-Hole Spectral Asymmetry at the Edge of Multi-orbital\\ Noncentrosymmetric Superconductors}

\author{Yuri Fukaya}
\affiliation{CNR-SPIN, I-84084 Fisciano (Salerno), Italy, c/o Universit\'a di Salerno, I-84084 Fisciano (Salerno), Italy}

\author{Keiji Yada}
\affiliation{Department of Applied Physics, Nagoya University, Nagoya 464-8603, Japan}

\author{Yukio Tanaka}
\affiliation{Department of Applied Physics, Nagoya University, Nagoya 464-8603, Japan}
\author{Paola Gentile}
\affiliation{CNR-SPIN, I-84084 Fisciano (Salerno), Italy, c/o Universit\'a di Salerno, I-84084 Fisciano (Salerno), Italy}

\author{Mario Cuoco}
\affiliation{CNR-SPIN, I-84084 Fisciano (Salerno), Italy, c/o Universit\'a di Salerno, I-84084 Fisciano (Salerno), Italy}

\begin{abstract}
\noindent 
In the Supplemental Material we show the methodology to obtain the surface Green's function. Then, we consider an effective two-orbital model with a pairing structure having orbital singlet symmetry and crystalline interactions that break the rotational symmetry. 
We show that symmetric surface density of states occurs in single-orbital noncentrosymmetric superconductor with Rashba coupling.
\end{abstract}
\maketitle

\section{Recursive Green's function method}

In this section we briefly present the recursive Green's function method that has been employed to calculate the surface density of states (SDOS) in a semi-infinite geometry.
The Hamiltonian is expressed by considering a real space representation along the $x$-axis (i.e.\ the direction of the edge termination), while the momentum is conserved along the $y$ direction. The Hamiltonian is hence given by
\begin{align}
    \tilde{H}^{total}(k_y)&=
    \begin{pmatrix}
        \tilde{u}(k_y) & \tilde{t}(k_y) & 0 & \cdots & 0 \\
        \tilde{t}^{\dagger}(k_y) & \tilde{u}(k_y) & \tilde{t}(k_y) & & \vdots \\
        0 & \tilde{t}^{\dagger}(k_y) & \tilde{u}(k_y) & \ddots & 0 \\
        \vdots & & \ddots & \ddots & \tilde{t}(k_y) \\
        0 & \cdots & 0 & \tilde{t}^{\dagger}(k_y) & \tilde{u}(k_y)
    \end{pmatrix}.
\end{align}%
Here, $\tilde{u}(k_y)$ denotes the local term in the superconducting state and $\tilde{t}(k_y)$ stands for the nearest neighbor hopping.
Then the surface Green's functions described by the recursive Green's function method~\cite{LDOSUmerski,KawaiPRB} are given by
\begin{align}
    \tilde{G}^\mathrm{L}_{N+1}(E,k_y)
    &=\frac{1}{z-\tilde{u}(k_y)-\tilde{t}^{\dagger}\tilde{G}^\mathrm{L}_{N}(E,k_y)\tilde{t}},\\
    \tilde{G}^\mathrm{R}_{N+1}(E,k_y)
    &=\frac{1}{z-\tilde{u}(k_y)-\tilde{t}\tilde{G}^\mathrm{R}_{N}(E,k_y)\tilde{t}^{\dagger}},
\end{align}%
with the retarded part $z=E+i\delta$, with the infinitesimal value $\delta$.
Here, we define the surface Green's function at $N=1$,
\begin{align}
    \tilde{G}^\mathrm{L}_{1}(E,k_y)&=\tilde{G}^\mathrm{R}_{1}(E,k_y)=\frac{1}{z-\tilde{u}(k_y)}.
\end{align}%
Repeating the above recursive Green's function method, we can obtain the surface Green's function at $N_\mathrm{SC}+1$ site with $N_\mathrm{SC}$ layers.

\section{Effective two-orbital model with orbital dependent pairings}

In this section we provide details of the effective two-orbital model assuming that the pairing potential is restricted to the equal spin sector (spin-triplet) and allowing for orbital singlet and orbital triplet configurations in the orbital sector. We determine the Green's function and show the explicit expression for its anomalous part in the presence of orbital splitting as due to reduced crystal symmetry at the edge.

The effective Hamiltonian for the two-orbital model in the $[as,bs]$ basis ($s=\uparrow,\downarrow$) is generally given by,
\begin{align}
    \hat{H}_\mathrm{BdG}(k)=
    \begin{pmatrix}
        \varepsilon_{a}(k) & 0 & \Delta_{aa}(k) & \Delta_{ab}(k) \\
        0 & \varepsilon_{b}(k) & \Delta_{ba}(k) & \Delta_{bb}(k) \\
        [\Delta_{aa}(k)]^{*} & [\Delta_{ba}(k)]^{*} & -\varepsilon_{a}(k) & 0 \\
        [\Delta_{ab}(k)]^{*} & [\Delta_{bb}(k)]^{*} & 0 & -\varepsilon_{b}(k)
    \end{pmatrix},
\end{align}%
with $\Delta_{\alpha\beta}\equiv\Delta^{T}_{\alpha s\beta s}$.
In the BdG Hamiltonian, the manifold of the orbital states, $(a,b)$ can be also represented through an effective pseudospin 1/2 algebra. 
We assume that the orbital states $(a,b)$ have the same parity with respect to the horizontal mirror transformation while an opposite parity with respect to the vertical mirror one (e.g. they indicate $xz$ and $yz$ orbitals). 
When considering the ($xz$,$yz$) manifold one can notice that there are different types of interactions that couple that two orbital configurations in a way to break either the vertical mirror or the rotational symmetry. Here, the $xz$ and $yz$ orbital states can make a basis for an effective pseudospin $1/2$ angular momentum, whose algebra is set by the usual $\eta_i (i=x,y,z)$ Pauli matrices. In this case, a term proportional to $\eta_z$ arises if the $C_4$ rotation symmetry, while terms proportional to $\eta_x$ or $\eta_y$ will break both the vertical mirror and the rotation symmetry. 
Then, without $C_4$ rotation symmetry as it occurs at the  (100) termination, one expects that the kinetic energy $\varepsilon_{a}(k)$ can be unequal to $\varepsilon_{b}(k)$.

Since we are restricting ourselves to the equal spin configuration (i.e. even upon exchange of the spin index), a general structure for the pair potential in the orbital sector is given by
\begin{align*}
    \tilde{\Delta}(\bm{k})&=[\tilde{\Psi}(\bm{k})+\bm{f}(\bm{k})\cdot\hat{\bm{\sigma}}]i\hat{\sigma}_{y}=
    \begin{pmatrix}
        \Delta_{aa}(\bm{k}) & \Delta_{ab}(\bm{k}) \\
        \Delta_{ba}(\bm{k}) & \Delta_{bb}(\bm{k})
    \end{pmatrix},
\end{align*}%
with the orbital-singlet/even-parity order parameter $\tilde{\Psi}(\bm{k})$ and orbital-triplet/odd-parity $\bm{f}(\bm{k})=[f_x(\bm{k}),f_y(\bm{k}), f_z(\bm{k})]$ defined as
\begin{align}
    \tilde{\Psi}(\bm{k})&=\frac{1}{2}[\Delta_{ab}(\bm{k})-\Delta_{ba}(\bm{k})],\\
    f_x(\bm{k})&=\frac{1}{2}[\Delta_{bb}(\bm{k})-\Delta_{aa}(\bm{k})],\\
    f_y(\bm{k})&=\frac{1}{2i}[\Delta_{aa}(\bm{k})+\Delta_{bb}(\bm{k})],\\
    f_z(\bm{k})&=\frac{1}{2}[\Delta_{ab}(\bm{k})+\Delta_{ba}(\bm{k})].
\end{align}%
Here, $\hat{\sigma}_{j=x,y,z,0}$ and $\hat{s}_{i=x,y,z,0}$ stand for the Pauli matrices in orbital and spin space, respectively.
In this framework, one can consider the pair potential for the orbital-singlet B$_{1g}$ and orbital-triplet B$_{2u}$ representations in the $D_{4h}$ point group:
\begin{align}
    \Delta_{aa}(\bm{k})&=\Delta_{T}[-\sin{k_y}+i\sin{k_x}],\\
    \Delta_{ab}(\bm{k})&=\Delta_{S}[\cos{k_x}-\cos{k_y}],\\
    \Delta_{ba}(\bm{k})&=-\Delta_{S}[\cos{k_x}-\cos{k_y}],\\
    \Delta_{bb}(\bm{k})&=\Delta_{T}[\sin{k_y}+i\sin{k_x}],
\end{align}%
for the orbital-singlet B$_{1g}$ $\Delta_{S}$ and orbital-triplet B$_{2u}$ pairings $\Delta_{T}$.

Here, the normal and anomalous parts of the retarded Green's function are given by
\begin{align}
    \hat{G}(E,\bm{k})&=
    \begin{pmatrix}
        G_{aa}(E,\bm{k}) & G_{ab}(E,\bm{k}) \\
        G_{ba}(E,\bm{k}) & G_{bb}(E,\bm{k})
    \end{pmatrix},\\
    \hat{F}(E,\bm{k})&=
    \begin{pmatrix}
        F_{aa}(E,\bm{k}) & F_{ab}(E,\bm{k}) \\
        F_{ba}(E,\bm{k}) & F_{bb}(E,\bm{k})
    \end{pmatrix},
\end{align}%
with $\bar{G}(E,\bm{k})=-\hat{G}^{*}(E,-\bm{k})$ and $\bar{F}(E,\bm{k})=\hat{F}^{*}(E,-\bm{k})$.
In this study, we can explicitly determine the expression of the Green function and it is given by
\begin{align}
    G_{aa}(E,\bm{k})&=\frac{[E+i\delta+\varepsilon_{a}(\bm{k})]\bar{Z}_{b}-[E+i\delta+\varepsilon_{b}(\bm{k})]|\Delta_{ba}(\bm{k})|^2}{\bar{Z}},\\
    G_{ab}(E,\bm{k})&=\frac{[E+i\delta+\varepsilon_{a}(\bm{k})]\Delta_{ab}(\bm{k})\Delta^{*}_{bb}(\bm{k})+[E+i\delta+\varepsilon_{b}(\bm{k})]\Delta_{aa}(\bm{k})\Delta^{*}_{ba}(\bm{k})}{\bar{Z}},\\
    G_{ba}(E,\bm{k})&=\frac{[E+i\delta+\varepsilon_{b}(\bm{k})]\Delta^{*}_{aa}(\bm{k})\Delta_{ba}(\bm{k})+[E+i\delta+\varepsilon_{a}(\bm{k})]\Delta^{*}_{ab}(\bm{k})\Delta_{bb}(\bm{k})}{\bar{Z}},\\
    G_{bb}(E,\bm{k})&=\frac{[E+i\delta+\varepsilon_{b}(\bm{k})]\bar{Z}_{a}-[E+i\delta+\varepsilon_{a}(\bm{k})]|\Delta_{ab}(\bm{k})|^2}{\bar{Z}},
\end{align}%
\begin{align}
    F_{aa}(E,\bm{k})&=\frac{\Delta_{aa}(\bm{k})\bar{Z}_{b}+\Delta_{ab}(\bm{k})\Delta_{ba}(\bm{k})\Delta^{*}_{bb}(\bm{k})}{\bar{Z}},\\
    F_{ab}(E,\bm{k})&=\frac{-\Delta_{ab}(\bm{k})|\Delta_{ba}(\bm{k})|^{2}+\Delta_{ab}(\bm{k})[E+i\delta+\varepsilon_{a}(\bm{k})][E+i\delta-\varepsilon_{b}(\bm{k})]+\Delta_{aa}(\bm{k})\Delta_{bb}(\bm{k})\Delta^{*}_{ba}(\bm{k})}{\bar{Z}},\\
    F_{ba}(E,\bm{k})&=\frac{-|\Delta_{ab}(\bm{k})|^{2}\Delta_{ba}(\bm{k})+\Delta_{ba}(\bm{k})[E+i\delta-\varepsilon_{a}(\bm{k})][E+i\delta+\varepsilon_{b}(\bm{k})]+\Delta_{aa}(\bm{k})\Delta_{bb}(\bm{k})\Delta^{*}_{ab}(\bm{k})}{\bar{Z}},\\
    F_{bb}(E,\bm{k})&=\frac{\Delta_{bb}(\bm{k})\bar{Z}_{a}+\Delta_{ab}(\bm{k})\Delta_{ba}(\bm{k})\Delta^{*}_{aa}(\bm{k})}{\bar{Z}},
\end{align}%
with
\begin{align}
    \bar{Z}&=\bar{Z}_{a}\bar{Z}_{b}+|\Delta_{ab}(\bm{k})|^2|\Delta_{ba}(\bm{k})|^2\notag\\
    &-[E+i\delta+\varepsilon_{a}(\bm{k})][E+i\delta-\varepsilon_{b}(\bm{k})]|\Delta_{ab}(\bm{k})|^2-[E+i\delta-\varepsilon_{a}(\bm{k})][E+i\delta+\varepsilon_{b}(\bm{k})]|\Delta_{ba}(\bm{k})|^2\notag\\
    &-\Delta_{aa}(\bm{k})\Delta_{bb}(\bm{k})\Delta^{*}_{ab}(\bm{k})\Delta^{*}_{ba}(\bm{k})-\Delta^{*}_{aa}(\bm{k})\Delta^{*}_{bb}(\bm{k})\Delta_{ab}(\bm{k})\Delta_{ba}(\bm{k}),
\end{align}%
\begin{align}
    \bar{Z}_{a}&=(E+i\delta)^2-\varepsilon^2_{a}(\bm{k})-|\Delta_{aa}(\bm{k})|^2,\\
    \bar{Z}_{b}&=(E+i\delta)^2-\varepsilon^2_{b}(\bm{k})-|\Delta_{bb}(\bm{k})|^2.
\end{align}%

We observe that the anomalous part of the Green's function has components with $F_{aa}(E,\bm{k})\neq F_{bb}(E,\bm{k})$ if $\Delta_{aa}\neq \Delta_{bb}$. Furthermore, the off-diagonal terms $F_{ab}(E,\bm{k})$ and $F_{ba}(E,\bm{k})$ are inequivalent if $\varepsilon_{a}(k)$ is different from $\varepsilon_{b}(k)$ as well as $\Delta_{ab} \neq \Delta_{ba}$. The energy splitting between the $a$ and $b$ states is also yielding an anomalous component of the Green's function with coexisting even and odd frequency terms.
Due to these considerations, one can observe that the structure of the anomalous Green's function at the edge can arise as a consequence of the presence of orbital singlet and orbital triplet pairing amplitudes. 
We recall that the orbital triplet configuration corresponds to an orbital quintet state if the $a$ and $b$ states belong to a manifold with $L=1$ angular momentum.


\section{Two-orbital model on finite size system}

In this section we construct an effective minimal model for a limited number of chains (i.e.\ two) in the 2D system. This is the minimal real space configuration that can include a pairing with orbital-singlet and spin-triplet pairing and allows us to introduce crystalline interactions that mimic the breaking of the rotational and mirror symmetry at the boundary of the superconductor. The model allows to have a spatial dependent pairing amplitude at the edge and the reduced crystalline point group symmetry. 
For this geometry, we just take two-orbital degrees of freedom as described in the Sect.\ II.
The BdG Hamiltonian for the two-orbital model, assuming an equal spin pairing, with two coupled superconducting chains along the $x$-direction is given by
\begin{align}
    \tilde{H}_\mathrm{BdG}(k_y)&=\hat{H}_{1,1}(k_y)+\hat{H}_{2,2}(k_y)+\hat{H}_{1,2},
\end{align}%
with
\begin{align}
    \hat{H}_{1,1}(k_y)&=\hat{H}_{2,2}(k_y)=
    \begin{pmatrix}
        \hat{H}_{0}(k_y) & \hat{\Delta}(k_y) \\
        \hat{\Delta}^{\dagger}(k_y) & -\hat{H}^{*}_{0}(-k_y)
    \end{pmatrix},\\
    \hat{H}_{1,2}&=
    \begin{pmatrix}
        \hat{t} & \hat{\Delta}_{t} \\
        \hat{\Delta}^{\dagger}_{t} & -\hat{t}^{*}
    \end{pmatrix},
\end{align}%
where $t$ refers to the nearest-neighbor chains.
Then, the Hamiltonian in the normal state for each layer is given by:
\begin{align}
    \hat{H}_{0}(k_y)&=
    \begin{pmatrix}
        -\mu-2t^{y}_{aa}\cos{k_y} & \varepsilon_{ab} \\
        \varepsilon^{*}_{ab} & -\mu-2t^{y}_{bb}\cos{k_y}
    \end{pmatrix},
\end{align}%
\begin{align}
    \hat{t}&=
    \begin{pmatrix}
        -t^{x}_{aa} & -t^{x}_{ab} \\
        -t^{x}_{ba} & -t^{x}_{bb}
    \end{pmatrix}
\end{align}%
For our purposes, we set the spin-triplet/orbital-singlet/$s$-wave pair potential for the intra-layer $\hat{\Delta}(k_y)$ and inter-layers $\hat{\Delta}_{t}$ components:
\begin{align}
    \hat{\Delta}(k_y)&=
    \begin{pmatrix}
        0 & |\Delta_{0}| \\
        -|\Delta_{0}| & 0
    \end{pmatrix},\\
    \hat{\Delta}_{t}&=0.
\end{align}%
This model can include the effects of the crystalline symmetry breaking interactions by suitably choosing the hopping amplitudes. 
For instance, assuming that $t^{y}_{aa} \neq t^{y}_{bb}$ with $t^{x}_{aa}=t^{x}_{bb}$ and $t^x_{ab}=0$ we can simulate a configuration that breaks the rotational symmetry. In a similar way, the presence of a nonvanishing term $t^x_{ab}$ will lead to both vertical mirror and rotational symmetry breaking.

\section{Orbital-splitting and rotation symmetry breaking at the edge: recursive Green's function approach}

In this section we analyze the role of the rotational symmetry breaking for a semi-infinite configuration by assuming that the system is split into two regions.
Indeed, we have a semi-infinite 2D superconductor and include five superconducting chains characterized by nonzero orbital splitting $\delta_{ab}$ at the (100) edge. The orbital splitting $\delta_{ab}$ transforms as $L^2_x -L^2_y$ in terms of the orbital angular momentum operators or equivalently as $\sigma_z$ in the ($xz,yz$) pseudospin subspace. As one can deduce from its operatorial form, it  breaks the $C_4$ rotational symmetry.
The regime with and without rotational symmetry would correspond to the bulk and the termination five-edge layers, respectively.
The outcomes are reported in Figure 1. One can see that the DOS at the edge is asymmetric in the presence of the orbital splitting at the edge.

\begin{figure}[htbp]
    \centering
    \includegraphics[width=8cm]{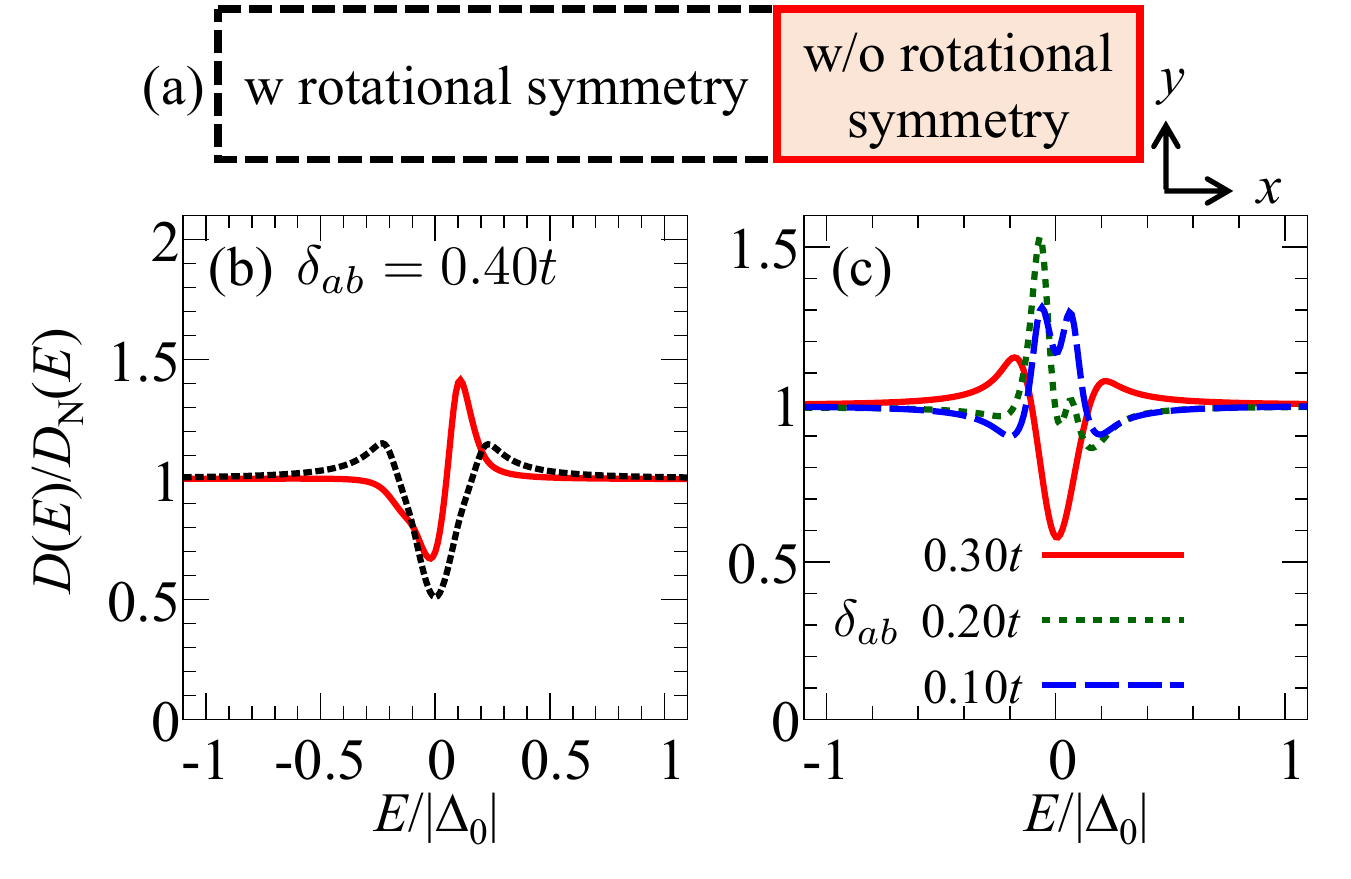}
    \caption{
    (a) Schematic illustration of a superconductor with and without $C_4$ rotational symmetry. 
    The regime with (without) rotational symmetry corresponds to the bulk (edge) region. 
    We consider a 2D semi-infinite superconductor and we consider that five superconducting chains are characterized by nonzero orbital splitting $\delta_{ab}$ at the edge that breaks $C_4$ rotation symmetry.
    (b) Density of states (DOS) in the bulk (black-dotted line) and surface DOS (red-solid) along $x$-direction at $\delta_{ab}=0.40\,t$.
    (c) Density of states in the region without $C_4$ rotational symmetry by changing $\delta_{ab}$.
    Due to the orbital-singlet and orbital-triplet components, asymmetric spectral weight at the edge appears.
    We fix the parameters as $\mu=0.35t$, $(\Delta_{S},\Delta_{T})=(0.90|\Delta_{0}|,0.10|\Delta_{0}|)$, and $|\Delta_{0}|=1.0\times 10^{-2}t$.}
    \label{SDOS_w_split_2D}
\end{figure}

The local and nearest-neighbor hopping matrices in this system are given by
\begin{align}
    \tilde{u}(k_y)&=
    \begin{pmatrix}
        \hat{u}(k_y) & \hat{u}_\mathrm{sc}(k_y) \\
        \hat{u}^{\dagger}_\mathrm{sc}(k_y) & -\hat{u}^{*}(-k_y)
    \end{pmatrix},\\
    \hat{u}(k_y)&=
    \begin{pmatrix}
        -\mu+2t_{1}+2t_{2}(1-\cos{k_y}) & 0 \\
        0 & -\mu+2t_{3}+t_{4}(1-\cos{k_y})
    \end{pmatrix},\\
    \hat{u}_\mathrm{sc}(k_y)&=
    \begin{pmatrix}
        -\Delta_{T}\sin{k_y} & -\Delta_{S}\cos{k_y}\\
        \Delta_{S}\cos{k_y} & \Delta_{T}\sin{k_y}
    \end{pmatrix},
\end{align}%
\begin{align}
    \tilde{t}(k_y)&=
    \begin{pmatrix}
        \hat{t}(k_y) & \hat{t}_\mathrm{sc}(k_y) \\
        \hat{t}^{\dagger}_\mathrm{sc}(k_y) & -\hat{t}^{*}(-k_y)
    \end{pmatrix},\\
    \hat{t}(k_y)&=
    \begin{pmatrix}
        -t_{1} & 0 \\
        0 & -t_{3}
    \end{pmatrix},\\
    \hat{t}_\mathrm{sc}(k_y)&=
    \begin{pmatrix}
        -\frac{\Delta_{T}}{2} & \frac{\Delta_{S}}{2}\\
        -\frac{\Delta_{S}}{2} & -\frac{\Delta_{T}}{2}
    \end{pmatrix}.
\end{align}%
In the edge few layers (we set five layers in this study), as shown in Fig.\ \ref{SDOS_w_split_2D} (a), we add the orbital orbital splitting $\delta_{ab}$ in the normal state, 
\begin{align}
    \hat{H}_{\delta ab}=\delta_{ab}\hat{\sigma}_{z}.
\end{align}%

\section{Hamiltonian structure for recursive method for three-orbital noncentrosymmetric superconductors}
 
In this section, we provide details o the local and nearest-neigbhor hopping matrices to calculate the SDOS for the three-orbital noncentrosymmetric superconductors. Here, 
$\tilde{u}(k_y)$ is given by
\begin{align}
    \tilde{u}(k_y)&=
    \begin{pmatrix}
        \hat{u}(k_y) & \hat{\Delta}_\mathrm{B1} \\
        \hat{\Delta}^{\dagger}_\mathrm{B1} & -\hat{u}^{*}(-k_y)
    \end{pmatrix}, 
\end{align}%
\noindent being
\begin{align}
    \hat{u}(k_y)&=\hat{u}_{0}(k_y)+\hat{H}_\mathrm{SO}+\hat{u}_\mathrm{is}(k_y),
\end{align}%
\noindent with 
\begin{align}
    \hat{u}_{0}(k_y)&=
    \begin{pmatrix}
        \tilde{\varepsilon}_{yz}(k_y) & 0 & 0 \\
        0 & \tilde{\varepsilon}_{zx}(k_y) & 0 \\
        0 & 0 & \tilde{\varepsilon}_{xy}(k_y)
    \end{pmatrix}\otimes\hat{s}_{0},\\
    \tilde{\varepsilon}_{yz}(k_y)&=2t_1(1-\cos{k_y})+2t_3,\\
    \tilde{\varepsilon}_{zx}(k_y)&=2t_1+2t_3(1-\cos{k_y}),\\
    \tilde{\varepsilon}_{xy}(k_y)&=4t_2-2t_2\cos{k_y}+\Delta_\mathrm{t},
\end{align}%
and
\begin{align}
    \hat{H}_\mathrm{SO}&=\lambda_\mathrm{SO}[\hat{L}_{x}\otimes\hat{s}_{x}+\hat{L}_{y}\otimes\hat{s}_{y}+\hat{L}_{z}\otimes\hat{s}_{z}],\\
    \hat{u}_\mathrm{is}(k_y)&=-\alpha_\mathrm{OR}\hat{L}_{x}\otimes\hat{s_{0}}\sin{k_y},
\end{align}%
where $t_1=t_2 (=t)$ and $t_3$ 
are the intraorbital hopping parameters, $\Delta_\mathrm{t}$
is the crystal field potential, and $\alpha_\mathrm{OR}$ the orbital Rashba (OR) coupling, respectively.
$\tilde{t}(k_y)$ indicates,
\begin{align}
    \tilde{t}(k_y)&=
    \begin{pmatrix}
        \hat{t}(k_y) & 0 \\
        0 & -\hat{t}^{*}(-k_y)
    \end{pmatrix},\\
    \hat{t}(k_y)&=
    \begin{pmatrix}
        -t_3 & 0 & \frac{\alpha_\mathrm{OR}}{2} \\
        0 & -t_1 & 0 \\
        -\frac{\alpha_\mathrm{OR}}{2} & 0 & -t_2
    \end{pmatrix}\otimes\hat{s_{0}}.
\end{align}%



 
\section{Anomalous part of the retarded Green's function at the edge}

\begin{figure*}[htbp]
    \centering
    \includegraphics[width=15cm]{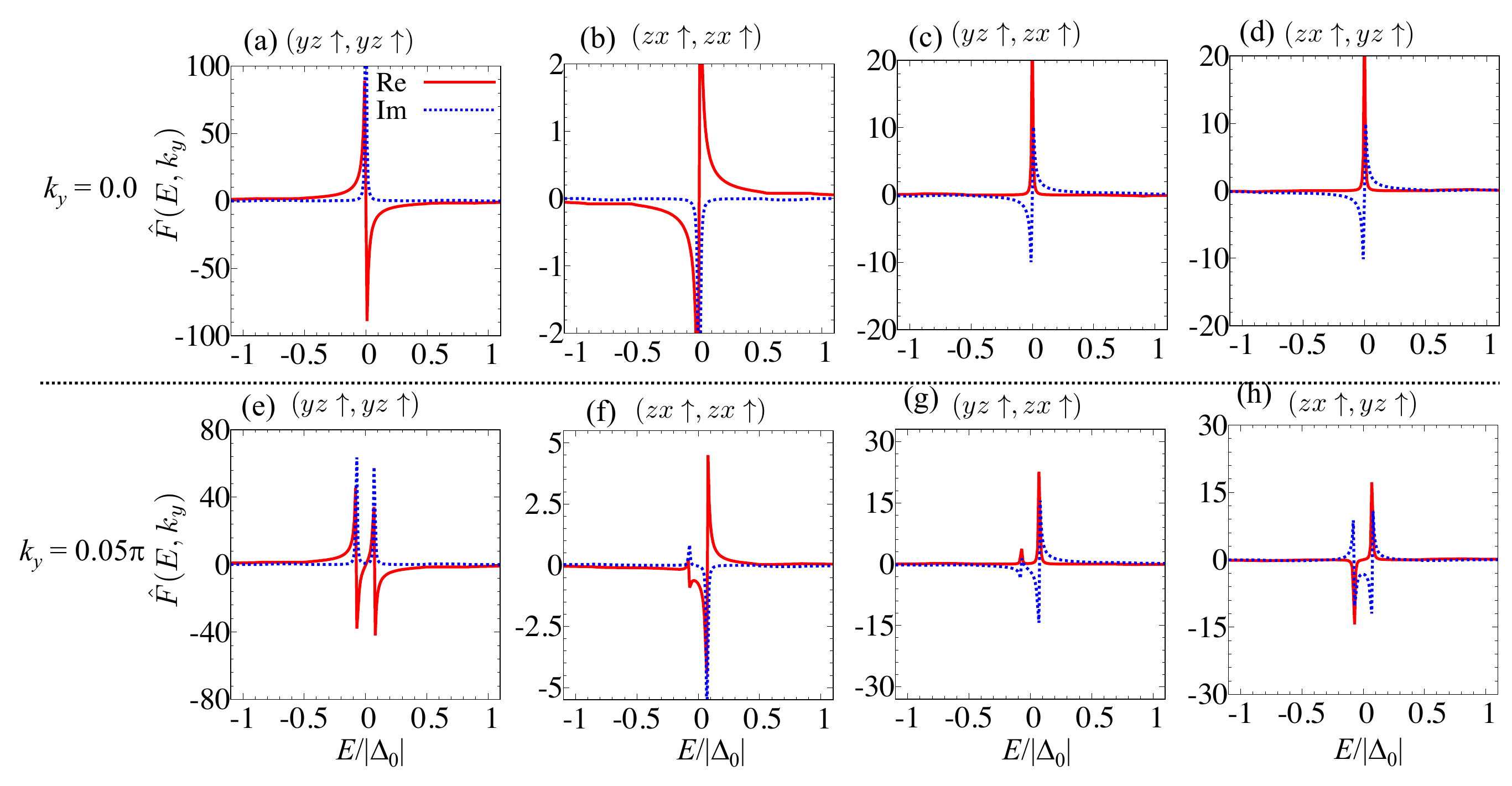}
    \caption{Anomalous part in the retarded Green's function $\hat{F}(E,k_y)$ for the spin-triplet/orbital-singlet/$s$-wave B$_1$ pairing at (a-d) $k_y=0$ and (e-h) $k_y=0.05\pi$ induced by OR coupling at the (100) edge for (a,e) $(yz\uparrow,yz\uparrow)$, (b,f) $(zx\uparrow,zx\uparrow)$, (c,g) $(yz\uparrow,zx\uparrow)$, and (d,h) $(zx\uparrow,yz\uparrow)$.
    Red-solid (blue-dotted) line denotes the real (imaginary) part of the anomalous part.
    We fix the OR coupling at $\alpha_\mathrm{OR}/t=0.20$, the chemical potential at $\mu/t=-0.25$, and the gap amplitude at $|\Delta_{0}|=0.0010t$.}
    \label{AGF_2FS}
\end{figure*}%
\begin{figure*}[htbp]
    \centering
    \includegraphics[width=14cm]{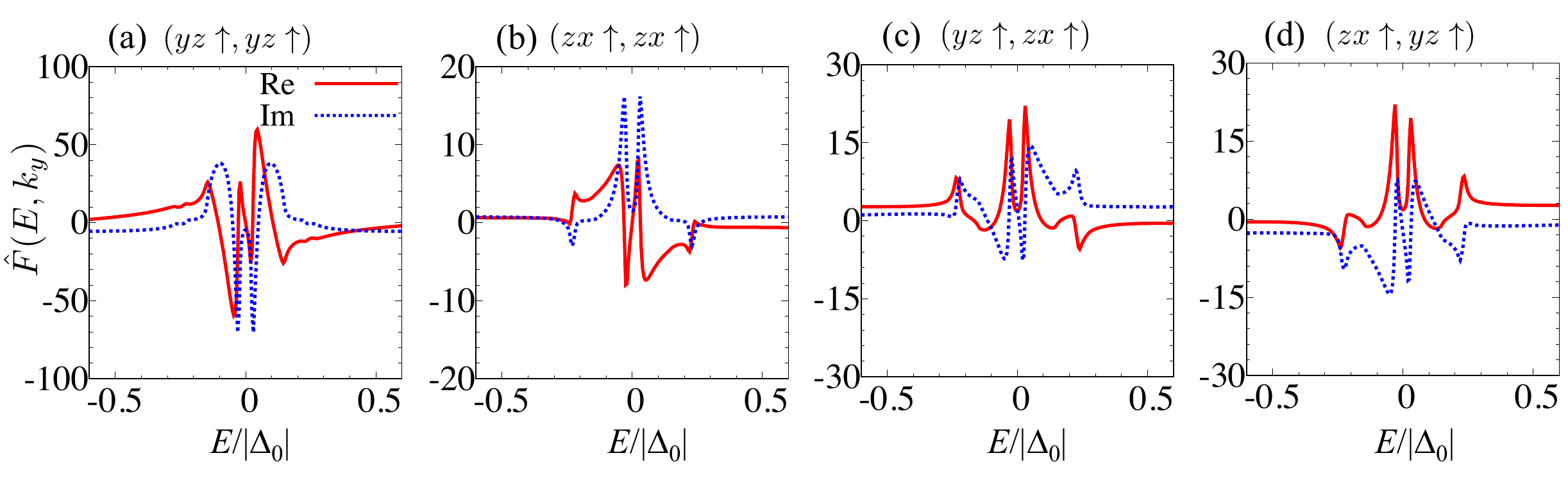}
    \caption{Anomalous part in the retarded Green's function $\hat{F}(E,k_y)$ for the spin-triplet/orbital-singlet/$s$-wave B$_1$ representation at $k_y=0$ induced by OR coupling at the (100) edge for (a) $(yz\uparrow,yz\uparrow)$, (b) $(zx\uparrow,zx\uparrow)$, (c) $(yz\uparrow,zx\uparrow)$, and (d) $(zx\uparrow,yz\uparrow)$.
    Red-solid (blue-dotted) line denotes the real (imaginary) part of the anomalous part.
    We fix the OR coupling at $\alpha_\mathrm{OR}/t=0.20$, the chemical potential at $\mu/t=0.35$, and the gap amplitude at $|\Delta_{0}|=0.0010t$.}
    \label{AGF_6FS}
\end{figure*}%
In this section we show the  anomalous part of the retarded Green's function induced by the orbital Rashba coupling at the (100) edge for the case of two Fermi surfaces (Fig.\ \ref{AGF_2FS}) and six Fermi surfaces at $k_y=0.0$ (Fig.\ \ref{AGF_6FS}).





\begin{figure}[htbp]
    \centering
    \includegraphics[width=13cm]{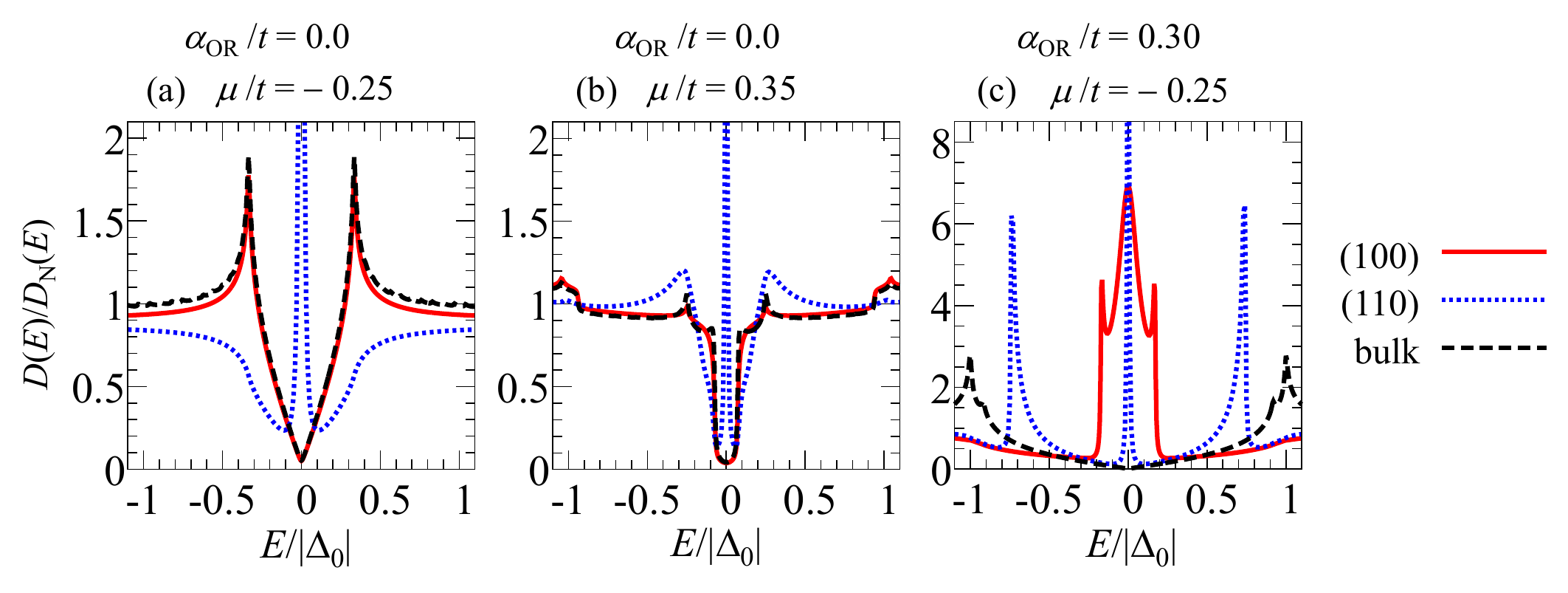}
    \caption{DOS for the spin-triplet/orbital-singlet/$s$-wave B$_1$ representation at (a,b) $\alpha_\mathrm{OR}/t=0.0$ and (c) at $\alpha_\mathrm{OR}/t=0.30$, respectively. 
    Then we choose the chemical potential as (a,c) $\mu/t=-0.25$ and (b) $\mu/t=0.35$.
    Red-solid, blue-dotted, and black-dotted lines denote the surface DOS (SDOS) at the (100) and (110) oriented surface, and DOS in the bulk, respectively.
    We fix the parameters as $\lambda_\mathrm{SO}/t=0.10$ and $|\Delta_{0}|/t=0.0010$.}
    \label{SMfig:4}
\end{figure}%
In the two-dimensional case, as shown in Fig.~\ref{SMfig:5} (a,b,c), the symmetric SDOS appears at both (100) and (110) oriented surfaces in the case with both two and six FSs at zero amplitude of the orbital Rashba coupling.

\section{Orbital-resolved SDOS}

\begin{figure*}[htbp]
    \centering
    \includegraphics[width=15cm]{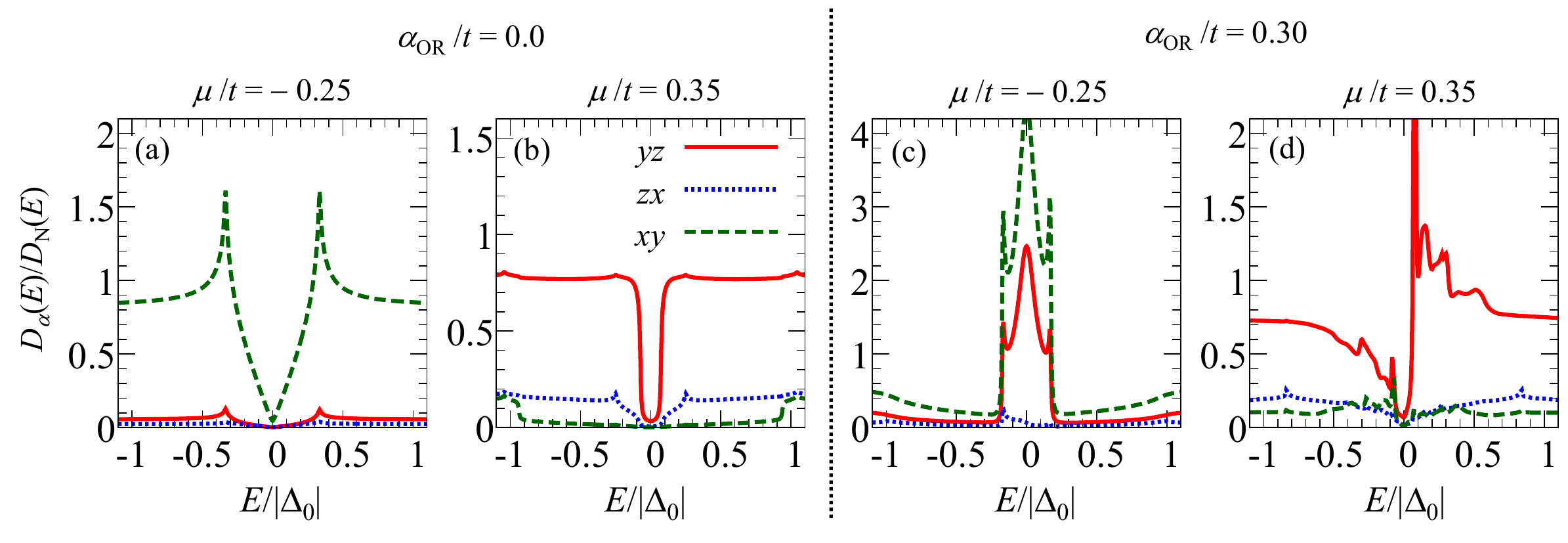}
    \caption{Orbital-resolved SDOS at the (100) oriented surface for the spin-triplet/orbital-singlet/$s$-wave B$_1$ representation at (a,b) $\alpha_\mathrm{OR}/t=0.0$ and (c,d) for $\alpha_\mathrm{OR}/t=0.30$, respectively. 
    Then we choose the chemical potential as (a,c) $\mu/t=-0.25$ and (b,d) $\mu/t=0.35$, respectively.
    Red solid, blue-dotted, and green-dotted lines denote the SDOS for $yz$, $zx$, and $xy$-orbitals, respectively.}
    \label{SMfig:5}
\end{figure*}%

In this section we discuss the orbital-resolved surface DOS for the spin-triplet/orbital-singlet/$s$-wave B$_1$ representation to explain which orbital configuration contributes to the SDOS at (100) oriented surface.
Orbital-resolved SDOS with orbital $\alpha=yz,zx,xy$ is defined as
\begin{align}
    D_{\alpha}(E)=-\frac{1}{\pi}\mathrm{Im}\int dk_y [G_{\alpha \uparrow,\alpha \uparrow}(E,k_y)+G_{\alpha \downarrow,\alpha \downarrow}(E,k_y)].
\end{align}%
In Fig.\ \ref{SMfig:5}, we plot the orbital resolved SDOS $D_{\alpha}(E)$ for $\alpha=yz$ (red-solid line), $zx$ (blue-dotted), and $xy$ (green-dotted) at (a,b) $\alpha_\mathrm{OR}/t=0.0$ and (c,d) $\alpha_\mathrm{OR}/t=0.30$.
Then we choose the chemical potential at (a,c) $\mu/t=-0.25$ (two FSs) and (b,d) $\mu/t=0.35$ (six FSs).
In the case with two FSs, the Hamiltonian is described by an effective single-orbital model and with Rashba coupling that is associated with $\alpha_\mathrm{OR}$.
As a result, each $D_{\alpha=yz,zx,xy}(E)$ is symmetric in Figs.\ \ref{SMfig:4} (a,c).
In the case of six FSs, SDOS for $yz$-orbital is mainly dominant and its asymmetry is enhanced by orbital Rashba $\alpha_\mathrm{OR}$ as shown in Figs.\ \ref{SMfig:5}(b,d).

\section{Asymmetric SDOS for the A$_1$ representation}


\begin{figure}
    \centering
    \includegraphics[width=15cm]{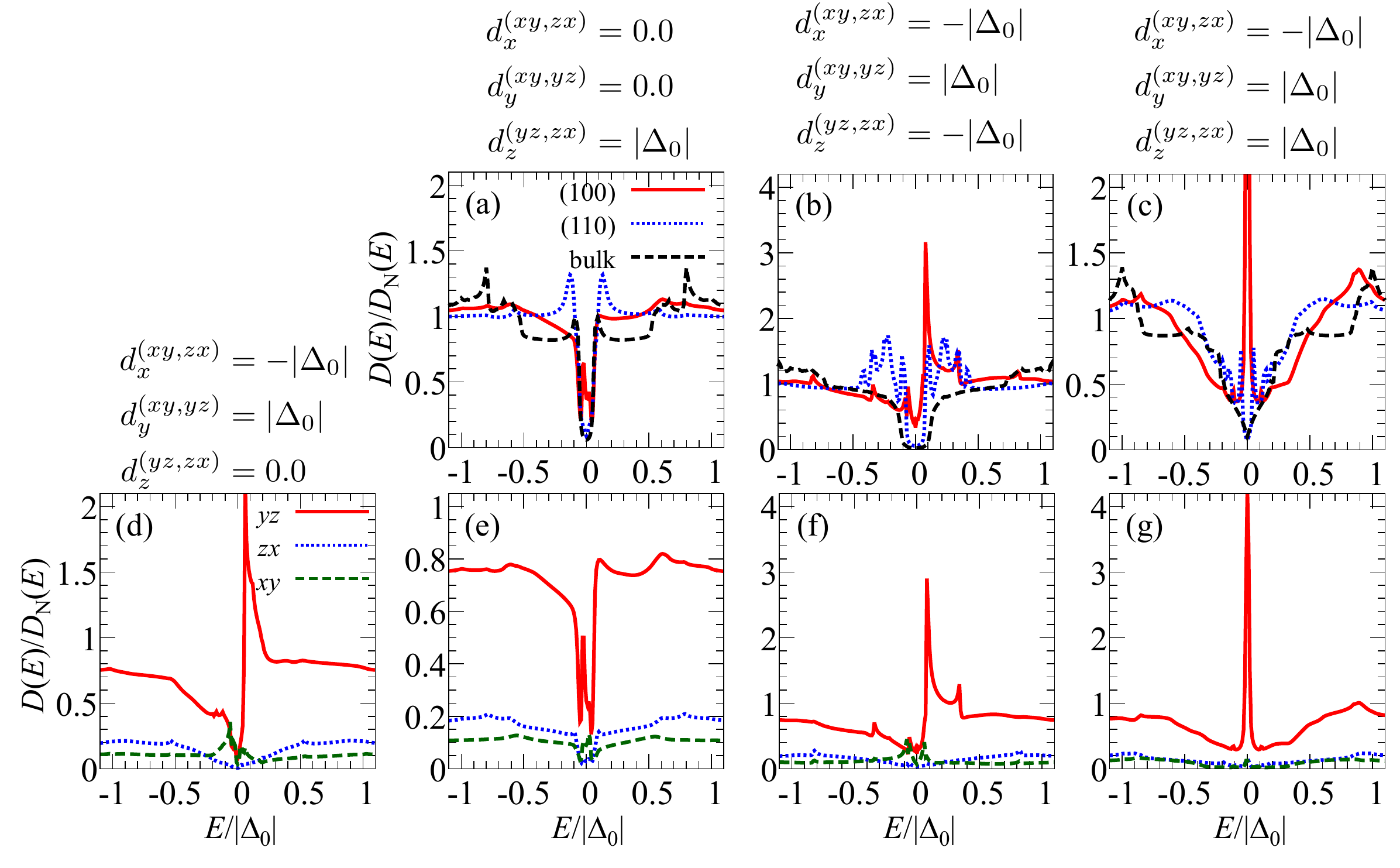}
    \caption{(a,b,c) DOS in the bulk (black-dotted line) and SDOS for the A$_1$ representation with a termination along the (100) (red-solid) and (110) (blue-dotted) direction. 
    (d,e,f,g) Orbital-resolved surface DOS for the A$_1$ pairing at the (100) oriented surface. 
    We set the pair potential as (a,e) $(d^{(xy,zx)}_{x},d^{(xy,yz)}_{y},d^{(yz,zx)}_{z})=(0,0,|\Delta_0|)$, (b,f) $(-|\Delta_0|,|\Delta_0|,-|\Delta_{0}|)$, (c,g) $(-|\Delta_0|,|\Delta_0|,|\Delta_{0}|)$, and (d) $(-|\Delta_0|,|\Delta_0|,0.0)$.
    We fix the parameters as $\mu/t=0.35t$, $\lambda_\mathrm{SO}=0.1t$, $\alpha_\mathrm{OR}=0.3t$, and $|\Delta_{0}|=0.001t$.}
    \label{SDOS_A1}
\end{figure}%
In this section we investigate the SDOS and orbital-resolved one for the A$_1$ pairing in the case with six Fermi surfaces at $\mu/t=0.35$ to show that asymmetric SDOS is also obtained.
The pair potential of the spin-triplet/orbital-singlet/$s$-wave A$_1$ representation \cite{fukaya18} is described by the orbital-dependent/momentum-independent $d$-vector: $d^{(xy,yz)}_{y}=-d^{(xy,zx)}_{x}$ and $d^{(yz,zx)}_{z}$ with
\begin{align}
    \hat{\Delta}_{\alpha,\beta}&=[\bm{d}^{(\alpha,\beta)}\cdot \hat{\bm{s}}]i\hat{s}_{y},\\
    \bm{d}^{(\alpha,\beta)}&=(d^{(\alpha,\beta)}_{x},d^{(\alpha,\beta)}_{y},d^{(\alpha,\beta)}_{z}).
\end{align}%
We choose the three kinds of the pair potential: (a,e) $(d^{(xy,zx)}_{x},d^{(xy,yz)}_{y},d^{(yz,zx)}_{z})=(0,0,|\Delta_0|)$, (b,f) $(-|\Delta_0|,|\Delta_0|,-|\Delta_{0}|)$, and (c,g) $(-|\Delta_0|,|\Delta_0|,|\Delta_{0}|)$, and (d) $(-|\Delta_0|,|\Delta_0|,0.0)$ as shown in Fig.\ \ref{SDOS_A1}.
As for the SDOS for the spin-triplet/orbital-singlet/$s$-wave B$_1$ pairing, asymmetry appears at the (100) oriented surface for the spin-triplet/orbital-singlet A$_1$ representation.
We note that the zero-energy peak of the SDOS along (100) direction originates from the accidental zero-energy Andreev bound states [Fig.\ \ref{SDOS_A1} (c)].

\section{SDOS for the case of single-orbital noncentrosymmetric superconductor}

In this section we analyze the SDOS for single-orbital noncentrosymmetric superconductors assuming various types of pairing configurations. This study aims to confirm that for single-orbital noncentrosymmetric superconductors the DOS at the edge is generally symmetric.
The Bogoliubov-de Gennes Hamiltonian for the single-orbital model with Rashba coupling $\Lambda$ is given by
\begin{align}
    \hat{H}_\mathrm{BdG}(\bm{k})&=
    \begin{pmatrix}
        \hat{H}(\bm{k}) & \hat{\Delta}(\bm{k}) \\
        \hat{\Delta}^{\dagger}(\bm{k}) & -\hat{H}^{*}(-\bm{k})
    \end{pmatrix},
\end{align}%
with the Hamiltonian in the normal state
\begin{align}
    \hat{H}(\bm{k})&=
    \begin{pmatrix}
        \varepsilon(\bm{k}) & \Lambda[\sin{k_y}+i\sin{k_x}] \\
        \Lambda[\sin{k_y}-i\sin{k_x}] & \varepsilon(\bm{k})
    \end{pmatrix},\\
    \varepsilon(\bm{k})&=-\mu+4t-2t\cos{k_x}-2t\cos{k_y},
\end{align}%
and the pair potential
\begin{align}
    \hat{\Delta}(\bm{k})&=
    \begin{cases}
        \psi(\bm{k})i\hat{s}_{y} & S=0 \\
        [\bm{d}(\bm{k})\cdot\hat{\bm{s}}]i\hat{s}_{y} & S=1
    \end{cases},
\end{align}%
for the spin-singlet $\psi(\bm{k})$ and the spin-triplet order parameter $\bm{d}(\bm{k})=(d_x(\bm{k}),d_y(\bm{k}),d_z(\bm{k}))$, respectively.
In the $C_{4v}$ point group, we set the pair potential as
\begin{align}
    \psi(\bm{k})&=\Delta_{S}(\cos{k_x}-\cos{k_y}),\\
    (d_x(\bm{k}),d_y(\bm{k}),d_z(\bm{k}))&=\Delta_{T}(\sin{k_y},\sin{k_x},0),
\end{align}%
for the B$_1$ representation and
\begin{align}
    \psi(\bm{k})&=\Delta_{S}\sin{k_x}\sin{k_y},\\
    (d_x(\bm{k}),d_y(\bm{k}),d_z(\bm{k}))&=\Delta_{T}(\sin{k_x},-\sin{k_y},0),
\end{align}%
for the B$_2$ one.

To calculate the SDOS at the (100) oriented surface, we obtain the local and nearest neighbor hopping terms,
\begin{align}
    \tilde{u}(k_y)&=
    \begin{pmatrix}
        \hat{u}(k_y) & \hat{u}_\mathrm{sc}(k_y) \\
        \hat{u}^{\dagger}_\mathrm{sc}(k_y) & -\hat{u}^{*}(-k_y)
    \end{pmatrix},\\
    \hat{u}(k_y)&=
    \begin{pmatrix}
        -\mu+4t-2t\cos{k_y} & \Lambda\sin{k_y} \\
        \Lambda\sin{k_y} & -\mu+4t-2t\cos{k_y}
    \end{pmatrix},
\end{align}%
and
\begin{align}
    \tilde{t}(k_y)&=
    \begin{pmatrix}
        \hat{t}(k_y) & \hat{t}_\mathrm{sc}(k_y) \\
        \hat{t}^{\dagger}_\mathrm{sc}(k_y) & -\hat{t}^{*}(-k_y)
    \end{pmatrix},\\
    \hat{t}(k_y)&=
    \begin{pmatrix}
        -t & -\frac{\Lambda}{2} \\
        \frac{\Lambda}{2} & -t
    \end{pmatrix},
\end{align}%
respectively.
Here, $\hat{u}_\mathrm{sc}(k_y)$ and $\hat{t}_\mathrm{sc}(k_y)$ are obtained by
\begin{align}
    \hat{u}_\mathrm{sc}(k_y)&=
    \begin{pmatrix}
        0 & -\Delta_{S}\cos{k_y}\\
        \Delta_{S}\cos{k_y} & 0
    \end{pmatrix},\\
    \hat{t}_\mathrm{sc}(k_y)&=
    \begin{pmatrix}
        0 & \frac{\Delta_{S}}{2}\\
        -\frac{\Delta_{S}}{2} & 0
    \end{pmatrix}
\end{align}%
for spin-singlet B$_1$,
\begin{align}
    \hat{u}_\mathrm{sc}(k_y)&=
    \begin{pmatrix}
        -\Delta_{T}\sin{k_y} & 0\\
        0 & \Delta_{T}\sin{k_y}
    \end{pmatrix},\\
    \hat{t}_\mathrm{sc}(k_y)&=
    \begin{pmatrix}
        -\frac{\Delta_{T}}{2} & 0 \\
        0 & -\frac{\Delta_{T}}{2}
    \end{pmatrix} 
\end{align}%
for spin-triplet B$_1$,
\begin{align}
    \hat{u}_\mathrm{sc}(k_y)&=0,\\
    \hat{t}_\mathrm{sc}(k_y)&=
    \begin{pmatrix}
        0 & i\frac{\Delta_{S}}{2}\sin{k_y}\\
        -i\frac{\Delta_{S}}{2}\sin{k_y} & 0
    \end{pmatrix},
\end{align}%
for spin-singlet B$_2$, and
\begin{align}
    \hat{u}_\mathrm{sc}(k_y)&=
    \begin{pmatrix}
        -i\Delta_{T}\sin{k_y} & 0 \\
        0 & -i\Delta_{T}\sin{k_y}
    \end{pmatrix},\\
    \hat{t}_\mathrm{sc}(k_y)&=
    \begin{pmatrix}
        -\frac{\Delta_{T}}{2} & 0 \\
        0 & \frac{\Delta_{T}}{2}
    \end{pmatrix},
\end{align}%
for spin-triplet B$_2$ representations in the $C_{4v}$ point group.

\begin{figure}[htbp]
    \centering
    \includegraphics[width=11cm]{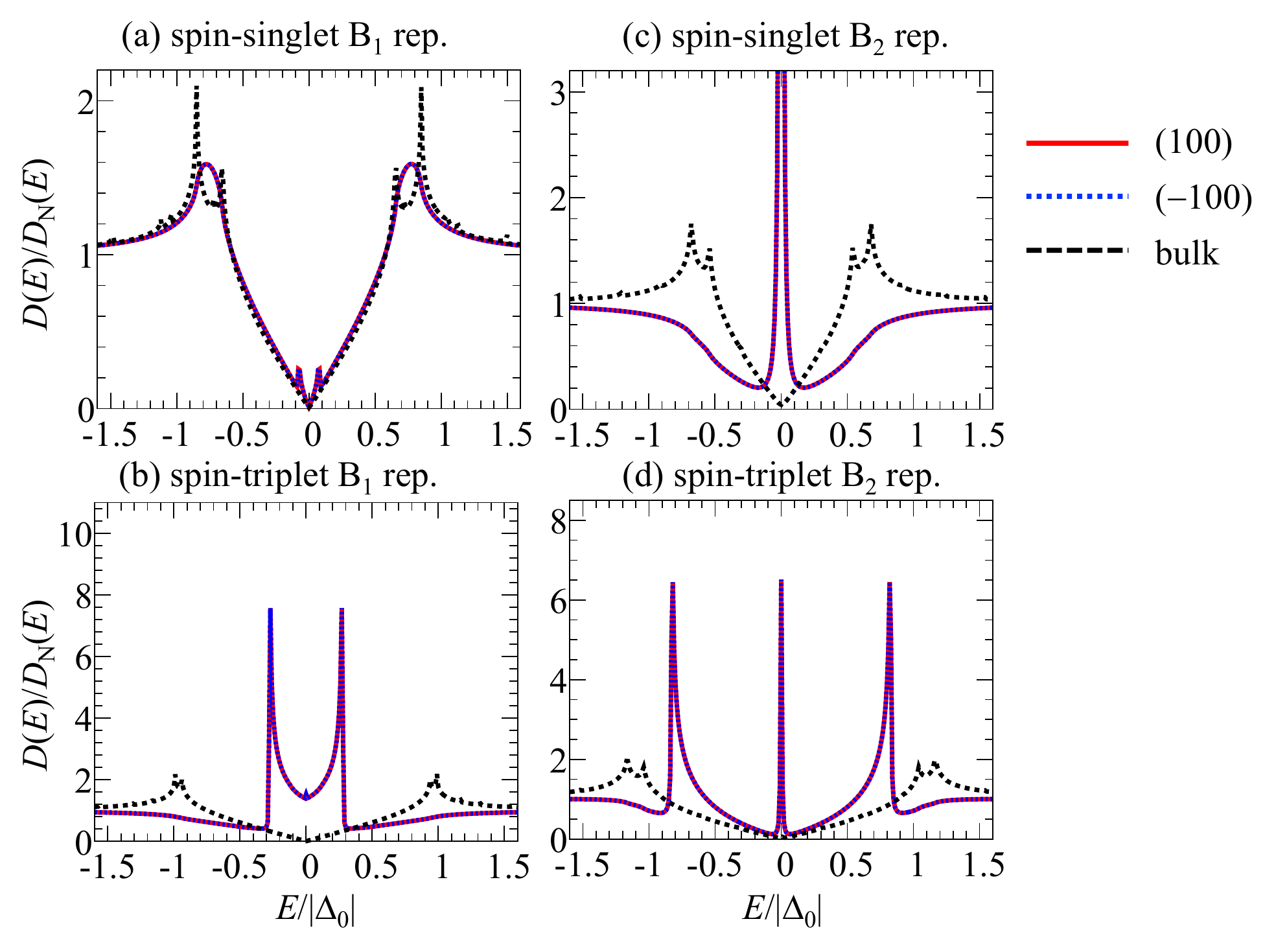}
    \caption{Surface density of states (SDOS) in the single-orbital model for (a) spin-singlet B$_1$, (b) spin-triplet B$_1$, (c) spin-singlet B$_2$, (d) spin-triplet B$_2$ representations.
    We set the parameters as $\mu/t=2.5$, $\Lambda=0.20t$, and $\Delta_{S}=\Delta_{T}=|\Delta_{0}|=0.001t$.}
    \label{SDOS_single_orbital}
\end{figure}%
We plot the DOS in the bulk (black-dotted line), and SDOS at the (100) (red solid) and $(-100)$ oriented surface (blue-dotted) in Fig.\ \ref{SDOS_single_orbital} for each irreducible representation.
All SDOS, as well as DOS in the bulk, is symmetric and they are independent of the Rashba coupling $\Lambda$.


%